\documentclass[sigconf]{acmart}
\AtBeginDocument{%
  }

\setcopyright{acmlicensed}
\copyrightyear{2025}
\acmYear{2025}
\acmDOI{XXXXXXX.XXXXXXX}
\acmConference[]{}{}

\usepackage{xcolor}
\usepackage{lipsum}
\usepackage{multirow}
\usepackage[most]{tcolorbox}
\tcbuselibrary{breakable} 
\tcbuselibrary{skins} 
\usepackage{subfigure}
\usepackage{tabularx}
\usepackage{bbm}
\usepackage{soul,color,xcolor}      
\definecolor{myColor}{rgb}{213, 0, 0}   
\newcommand{\review}[1]{\textcolor{black}{#1}}  

\usepackage[textsize=tiny]{todonotes}

\begin{document}

\title[AutoTestForge]{AutoTestForge: A Multidimensional Automated Testing Framework for Natural Language Processing Models}

\author{Hengrui Xing}
\email{HruiX@stu.xidian.edu.cn}
\affiliation{%
	\institution{Xidian University}
	\city{Xi'an}
	\state{ShaanXi}
	\country{China}
}

\author{Cong Tian}
\email{ctian@mail.xidian.edu.cn}
\affiliation{%
	\institution{Xidian University}
	\city{Xi'an}
	\state{ShaanXi}
	\country{China}
}

\author{Liang Zhao}
\email{lzhao@xidian.edu.cn}
\affiliation{%
	\institution{Xidian University}
	\city{Xi'an}
	\state{ShaanXi}
	\country{China}
}

\author{Zhi Ma}
\email{mazhi@xidian.edu.cn}
\affiliation{%
	\institution{Xidian University}
	\city{Xi'an}
	\state{ShaanXi}
	\country{China}
}

\author{Wensheng Wang}
\email{wswang@xidian.edu.cn}
\affiliation{%
	\institution{Xidian University}
	\city{Xi'an}
	\state{ShaanXi}
	\country{China}
}

\author{Nan Zhang}
\email{nanzhang@xidian.edu.cn}
\affiliation{%
	\institution{Xidian University}
	\city{Xi'an}
	\state{ShaanXi}
	\country{China}
}

\author{Chao Huang}
\email{chao.huang@soton.ac.uk}
\affiliation{%
	\institution{University of Southampton}
	\city{Southampton}
	\state{Hampshire}
	\country{United Kingdom}
}
\author{Zhenhua Duan}
\email{zhhduan@xidian.edu.cn}
\affiliation{%
	\institution{Xidian University}
	\city{Xi'an}
	\state{ShaanXi}
	\country{China}
}

\begin{abstract}
	In recent years, the application of behavioral testing in Natural Language Processing (NLP) model evaluation has experienced a remarkable and substantial growth. However, the existing methods continue to be restricted by the requirements for manual labor and the limited scope of capability assessment. To address these limitations, we introduce AutoTestForge, an automated and multidimensional testing framework for NLP models in this paper. Within AutoTestForge, through the utilization of Large Language Models (LLMs) to automatically generate test templates and instantiate them, manual involvement is significantly reduced. Additionally, a mechanism for the validation of test case labels based on differential testing is implemented which makes use of a multi-model voting system to guarantee the quality of test cases. The framework also extends the test suite across three dimensions, taxonomy, fairness, and robustness, offering a comprehensive evaluation of the capabilities of NLP models. This expansion enables a more in-depth and thorough assessment of the models, providing valuable insights into their strengths and weaknesses. A comprehensive evaluation across sentiment analysis (SA) and semantic textual similarity (STS) tasks demonstrates that AutoTestForge consistently outperforms existing datasets and testing tools, achieving higher error detection rates (an average of $30.89\%$ for SA and $34.58\%$ for STS). Moreover, different generation strategies exhibit stable effectiveness, with error detection rates ranging from $29.03\% - 36.82\%$.
\end{abstract}





\maketitle
\section{Introduction}

\begin{figure*}[h]
	\centering
	\includegraphics[width=0.9\textwidth]{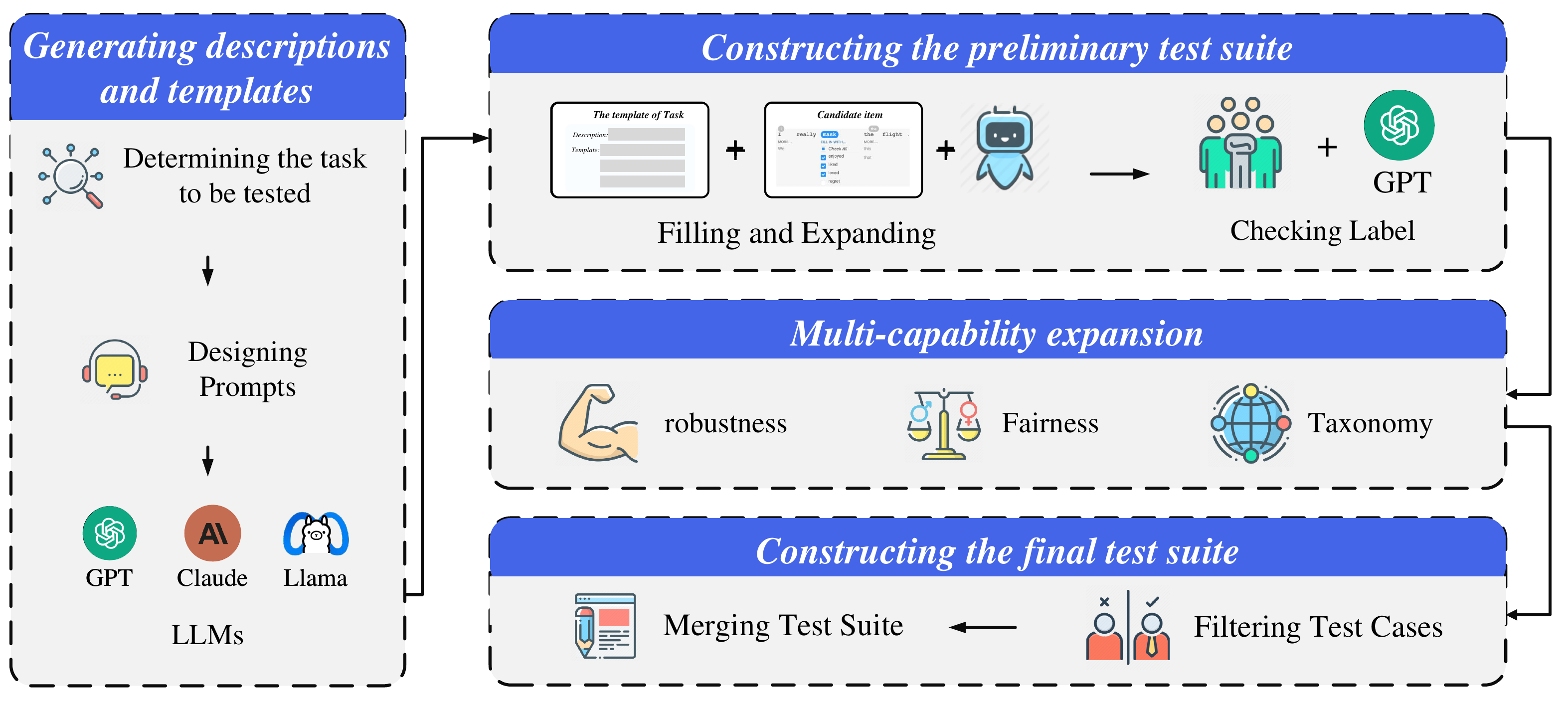}
	\caption{Overview of AutoTestForge}
	\label{fig:AutoTestForge}
\end{figure*}


Natural language processing (NLP) models have demonstrated excellent performance in analyzing and handling natural language, attracting widespread attention and significantly influencing various aspects of our lives \cite{khuranaNaturalLanguageProcessing2023,sefaraToolkitTextExtraction2022}.
However, it has been discovered that these models may be vulnerable to adversarial attacks, or affected by biases and noise present in the training data, potentially leading to unexpected outputs. This can result in misunderstandings \cite{okrentLittleTranslationMistakes2016}, economic losses \cite{economiclossesnews}, threats to personal safety \cite{xiangHeWouldStill2023}, and even political conflicts \cite{doi:10.1126/science.aau2706}.
Consequently, it is essential to conduct comprehensive test evaluation on NLP systems, which should cover a wide range of possible risks, including adversarial attacks \cite{23aseleap}, inherent biases \cite{XiaoFairnessTesting2023}, and other relevant concerns \cite{tan-etal-2021-reliability}, to identify and mitigate potential safety hazards.




The standard practice for model evaluation is to divide data into training, validation, and test sets. Specifically, the training set serves as the basis for model learning, the validation set is utilized for fine-tuning the model, and the test set is employed for the final assessment of model performance, as detailed in \cite{books/sp/Zhou21a}.
Within the research community, it is a common practice to submit model performance results to public evaluation platforms (leaderboards) and engage in comparisons. This practice enables the tracking of technological advancements in the field, as demonstrated in \cite{wang-etal-2018-glue, kiela-etal-2021-dynabench}.
However, this evaluation method is fraught with limitations. Test data frequently emanates from sources closely resembling those of the training data, and as a result, may harbor identical data biases. This situation has the potential to lead to an overestimation of model performance, as elucidated in \cite{pmlr-v97-recht19a, belinkov2018synthetic}.
Furthermore, relying on single performance metrics presents another challenge. It renders the identification of specific model deficiencies arduous and fails to offer clear-cut guidance for model enhancement, as indicated by \cite{wu-etal-2019-errudite, ribeiroAccuracyBehavioraltesting2020}.

In order to conduct more rigorous evaluations of NLP systems, researchers have put forward a variety of innovative evaluation methods.
Among these methods, fairness evaluation aims to determine whether the model's performance across different demographic groups, such as race and gender, is biased \cite{XiaoFairnessTesting2023}. Robustness evaluation is concerned with assessing the stability of a model's prediction results when faced with adversarial attacks and input perturbations \cite{23aseleap}. Additionally, vocabulary ability evaluation focuses on the model's overall performance in terms of its lexical coverage and the depth of its semantic understanding \cite{pilehvar-camacho-collados-2019-wic}.
Most of the existing methods examine the model from a wide variety of perspectives. They uncover issues that are not easily detected when relying solely on aggregate indicators. However, these methods typically focus on specific tasks or capabilities, falling short of comprehensively evaluating the model.



To address this issue, the Checklist proposed by Ribeiro et al. \cite{ribeiroAccuracyBehavioraltesting2020} divides the input space into different language phenomena and test types, and designs corresponding test cases for each phenomenon and type.
While this tool is effective in comprehensively evaluating NLP models, it still has some significant drawbacks. 
{Specifically, designing high quality test suites demands substantial expertise and creativity, which mainly relies on manual completion \cite{bhattCasestudyefficacy2021}.}
{An annotator typically requires approximately one hour to generate $5-7$ templates covering only $1-2$ capabilities, with a substantial time investment of $0.5-2$ working days necessary for producing a comprehensive set of multi-dimensional test templates.
	To improve efficiency,  recent studies}
\cite{ribeiroAdaptiveTestingDebugging2022,rastogiInvestigatingrelativestrengths2023,yangTestAugFrameworkAugmenting2022} have explored the possibility of automated test generation using generative models in conjunction with human intervention.
Nevertheless, the process of creating templates or designing test topics still requires substantial human involvement, which significantly constrains the efficiency and scalability of test creation.
{Further, individual test cases often evaluate isolated model capabilities, overlooking the fact that real-world tasks typically demand multiple competencies simultaneously.
	Failure to account for the intricate connections between various capabilities may lead to overestimating the model's actual performance.}
In summary, the existing NLP evaluation techniques still face challenges in reducing human dependency and conducting multidimensional capability assessment.


In light of these challenges, this paper puts forward AutoTestForge, an automated and multi-dimensional testing framework designed for NLP models. This framework serves to diminish the reliance on domain experts while comprehensively assessing the performance of NLP systems across a broad array of capabilities.
As depicted in Figure \ref{fig:AutoTestForge}, the process commences with the identification of the target task for evaluation. Subsequently, large language models (LLMs) are harnessed to generate detailed sentence structure descriptions and corresponding templates. These templates are then expanded via filling-expansion language models, giving rise to a rich variety of test cases. To guarantee the precision of the generated test cases, differential testing is adopted to authenticate the labels of test cases. Thereafter, a voting mechanism is implemented to single out the most dependable ones, and those judged to be inaccurate are rectified.
Following this stage, the generated test suite undergoes further refinement to augment vocabulary coverage, fairness, and robustness. This iterative refinement process meticulously incorporates all the test cases yielded through the preceding steps, culminating in a final, high-quality test suite that facilitates comprehensive and exacting evaluation.



To validate the effectiveness and practicality of our approach, we have implemented AutoTestForge \footnote{\url{https://figshare.com/s/b8c0c1d2e9224fd1eba7}} and carried out extensive experiments across sentiment analysis and semantic textual similarity tasks. 
Our experimental evaluation concentrates on three crucial aspects: the overall performance of the framework, the quality of test templates generated by LLMs, and the effectiveness of the resulting test cases.
The experimental results reveal the consistent advantages of AutoTestForge across various testing scenarios. When compared with traditional datasets and state-of-the-art  tools, AutoTestForge attains superior error detection rates, with {significant improvements in both sentiment analysis and semantic textual similarity tasks}. Moreover, our framework is effectively capable of reducing performance disparities between templates generated by different LLMs.

The major contributions made by this research are encapsulated as follows:
\begin{itemize}
	\item We propose a fully automated {test case} generation framework named AutoTestForge, which harnesses the power of LLMs to autonomously generate test templates, thereby eliminating the need for manual design;
	
	\item AutoTestForge seamlessly integrates multiple testing dimensions, including vocabulary assessment, fairness evaluation, and robustness analysis, into a unified and comprehensive framework; and
	
	\item experimental evaluations have shown that AutoTestForge attains state-of-the-art performance. Specifically, it demonstrates an edge in the average error detection rate, outperforming existing testing tools and traditional datasets by $10\%-20\%$.	
\end{itemize}

The remainder of this paper is structured as follows. Section \ref{sec:Back} provides an overview of the relevant background knowledge. In Section \ref{sec:framework}, we introduce AutoTestForge, the proposed automated and multi-dimensional testing framework.
 Section \ref{sec:exp} implements the framework and conducts a series of comparative experiments to evaluate its performance.
Section \ref{sec:related} reviews the related work. Finally, Section \ref{sec:con} summarizes the paper and discusses the potential future work.

\section{BACKGROUND CONCEPTS} \label{sec:Back}

Before presenting the framework, we introduce relevant background concepts, such as behavioral testing and test templates.

\subsection{Behavioral Testing}

Testing plays a pivotal role in safeguarding the quality and reliability of software products. Its primary objective is to detect defects or errors that could potentially lead to incorrect system behavior, as noted in \cite{wangSoftwaretestinglarge2024}. Generally speaking, software testing can be classified into two main categories: white-box testing and black-box testing.
Black-box testing, alternatively referred to as behavioral testing, focuses on validating the input-output behavior of software without the need for insights into its internal structure.

{In recent years, the concept of behavioral testing has been introduced in the NLP field~\cite{ribeiroAccuracyBehavioraltesting2020, yangTestAugFrameworkAugmenting2022,hlavnovaEmpoweringCrosslingualBehavioral2023,vanakenWhatYouSee2022a,wuGeneralErrorDiagnosis2023}. Researchers evaluate model capabilities through designing diverse test scenarios, with these capabilities being systematically categorized based on linguistic theory, logical reasoning, domain knowledge, or specific application requirements. For different NLP tasks, we can design targeted test cases to verify specific model capabilities. For example, in SA tasks, we can use examples like "\textit{I don't like this movie.}" to test whether the model can accurately interpret input sentences that contain positive words in combination with negation. In STS tasks, we can use question pairs like "\textit{Is Mary a teacher?}" and "\textit{Is Mary an accredited teacher?}" to test whether the model can understand how modifiers affect the semantic meaning of questions.}

{The identification and classification of these testing capabilities typically rely on the experience of software developers or systematic organization by domain experts with professional knowledge in data annotation. 
However, these methods require substantial human resources and time. }
Moreover, even experienced experts struggle to exhaustively enumerate all possible test scenarios, resulting in insufficient coverage. 
These limitations hinder a comprehensive and effective assessment of the quality and reliability of NLP systems.

\subsection{Testing Templates}
\begin{figure}[h]
	\centering
	\includegraphics[width=\linewidth]{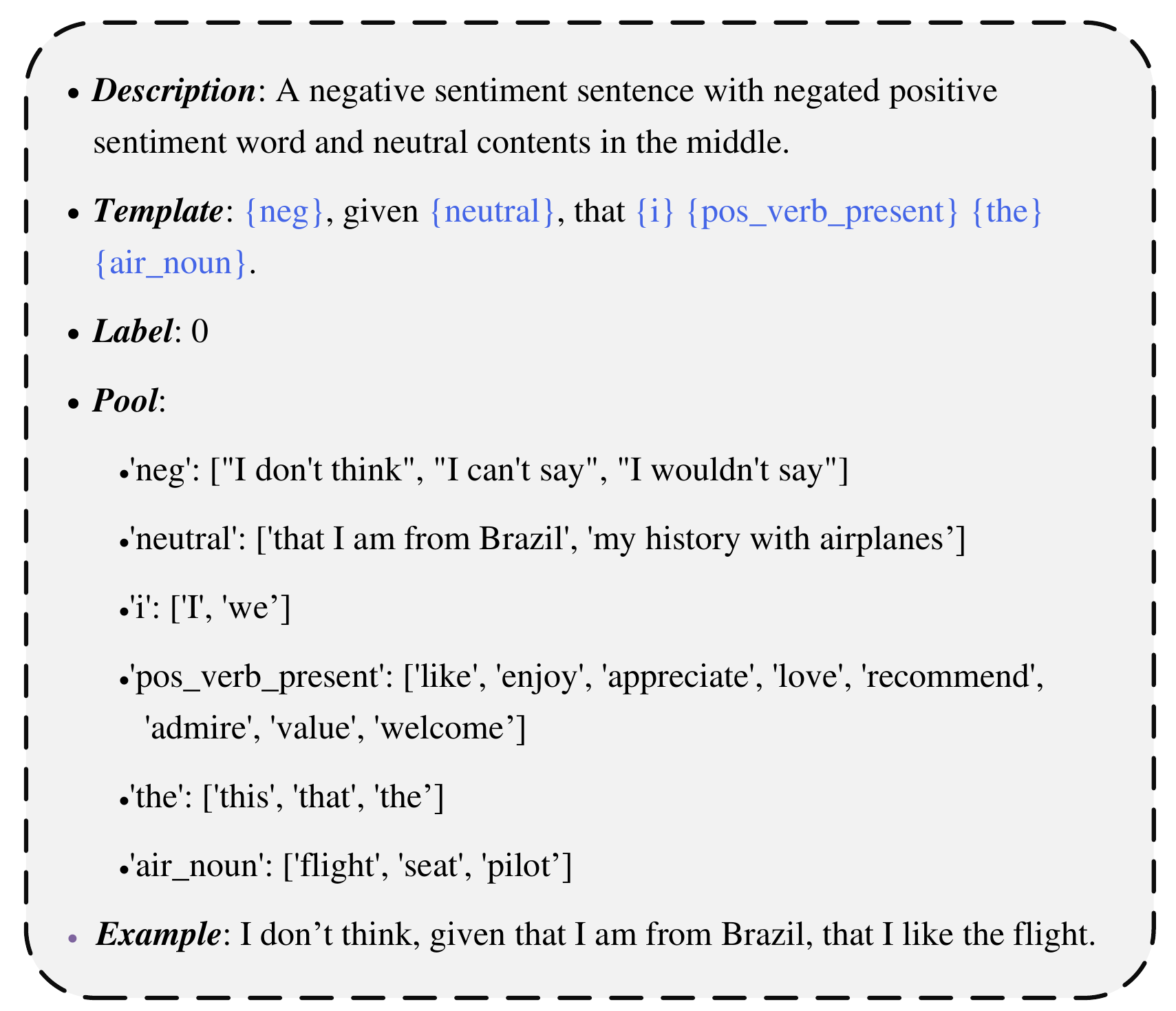}
	\caption{An example of test template}
	\label{fig:example-template}
\end{figure}

In the realm of behavioral testing, test case generation stands as a pivotal step that significantly influences the comprehensiveness and efficiency of the testing process. A test case can be delicately crafted based on a well-designed test template, complemented by corresponding lists of candidate words. This structured approach ensures that test case generation remains flexible and controllable. To gain a deeper understanding of this process, it is necessary to define several key concepts. An illustrative example of these concepts is provided in Figure \ref{fig:example-template}.

\textit{Description} refers to the sentence structure. It provides a fundamental framework to ensure grammatical and structural accuracy. \textit{Template}, an evolution of the structure, is a word-filling schema containing one or more slots, each linked to a predefined vocabulary list. This design offers flexibility in generating diverse test sentences through word substitution. \textit{Label} defines the classification categories for the task, typically represented numerically (e.g., $0, 1$). It represents the model's output categories. Actually, a test template is essentially a sentence structure with placeholders that can be populated with words from candidate lists.

Developers can leverage these templates to generate a diverse array of test cases tailored to specific testing requirements. These individual test cases are then aggregated into a comprehensive test suite. However, current template-based approaches to behavioral testing of NLP models face a significant challenge in efficiency: each test case is typically designed and utilized to assess a single capability separately. The one-to-one correspondence between cases and capabilities can be formally represented as a function. Let $T$ denote the test suite and $C$ represent the set of capabilities to be evaluated. The bijective mapping between the test suite and capabilities can be expressed as $f: T \to C$, where for each test case $t \in T$ in test suite, there exists a unique capability $c \in C$ such that $f(t) = c$.
While this approach is intuitively appealing, it gives rise to several efficiency concerns, including suboptimal resource utilization, constrained test capability coverage, and difficulty in capturing the intricacy among various capabilities.


\section{The AutoTestForge Framework} \label{sec:framework}

This section discusses the proposed testing framework AutoTestForge in detail.
It aims to significantly reduce repetitive manual testing efforts through automated testing techniques. In addition, it strives to maximize the potential value of each test case, comprehensively expanding the breadth and depth of testing.
Figure \ref{fig:AutoTestForge} shows the overall process of the AutoTestForge framework.
It incorporates four integral phases: generating descriptions and templates, constructing the preliminary test suite, muti-capability expansion, and constructing the final test suite. These phases operate in a synergistic manner, constituting an efficient testing workflow.

\subsection{Generating Descriptions and Templates} \label{sec:framework-A}

To enhance LLM-generated descriptions and templates, we have developed a structured prompting framework comprising four interconnected components, including background setting, definition declaration, special guidance, and output specifications.

The \textit{background setting} component provides LLMs with task-specific context, such as the nature of the current NLP task, its objectives, and potential challenges. This comprehensive information helps LLMs better understand the task requirements, resulting in more relevant and focused descriptions and templates. Specifically, background setting is provided as follows.

\begin{tcolorbox}[colback=gray!10,colframe=black,arc=2mm, auto outer arc,title={\textbf{Background Setting}},breakable,]
	As a linguist, your expertise is in modifying sentence structures and analyzing \verb|<task>|.
	
	You can construct completely different sentence structures based on different \verb|<task>| tasks. Each sentence structure will be unique and highly representative.
	
	And you will generate relevant content in \verb|<task_scenario>| scenario.
\end{tcolorbox}

The \textit{definition declaration} component outlines clear definitions for key terms and concepts, including the description, template, candidate words (pool), label, and example. We establish these definitions as follows, so as to reduce ambiguity and improve the accuracy and consistency of LLMs' outputs.
a) Description: the syntactic structure of a sentence. When generating templates based on a description, LLMs should adhere to this specified structure.
b) Template: a word-filling template derived from the sentence structure. Candidate words within the template are enclosed in curly braces ``\verb|{}|". The set of candidate words constitute the pool.
c) Label: the current task incorporates \verb|x| labels, denoted as \verb|0-x| and \verb|1-x|.
d) Pool: a collection of words intended to fill the placeholders (marked by ``\verb|{}|") in the template.
e) Example: a complete sentence obtained by populating the template with words from the pool.
The purpose of including an example is to ensure language models can
verify whether filling the template with words from the output pool yields coherent and fluent sentences.

The \textit{special guidance} component incorporates detailed instructions and exemplars, which are designed to steer language models toward generating more fluent and task-appropriate descriptions and templates.
A better specific guidance should be achieved by providing multifaceted directives within the prompt given to the language models as follows.
a) To ensure the generated template structures align with the desired objectives, the target task and the content of the required task labels are consistently emphasized throughout the prompt.
b) To obtain more fluent target templates, the language models are required to score the fluency and naturalness of the template. Only templates scoring 9.5 or above (out of 10) are returned.
c) To enhance the templates' ability to identify a broader range of defects, specific requirements are incorporated into the prompt.
d) Relevant criteria are established for the scenarios that language models generate templates, ensuring contextual appropriateness.
This comprehensive approach aims to optimize the performance of language models in generating high-quality, task-specific templates and descriptions.

Finally, the \textit{output specification} component specifies the model's results into a structured form for subsequent processing. It is stipulated that the language models should return their output in the JSON format. The format template is provided as follows.

\begin{tcolorbox}[colback=gray!10,colframe=black,arc=2mm, auto outer arc,title={\textbf{Output} },breakable,]
	\{\\
	``Description": \verb|<term>|, \\
	``Templates":[ \\
	\{\\
	\indent\quad``template":\verb|<term>|,	\quad	``label":\verb|<term>|, \\
	\indent\quad``pool":\verb|<term>|,		\qquad \enspace	``example": \verb|<term>|, \\
	\indent\quad``check\_label":\verb|<term>|,\quad \enspace	``score":\verb|<term>| \\
	\},\\
	{\verb|<term>|},***] \\
	\}
\end{tcolorbox}

Throughout the entire process, the four components work in synergy, guiding and controlling the generation of high-quality, diverse, and targeted sentence structure descriptions and templates by LLMs through refined prompt design.
The structured prompt strategy is highly extensible and adaptable. As new linguistic phenomena emerge or testing requirements evolve, the modular components of this framework can be readily adjusted and extended.  {An example of a prompt for a sentiment analysis task is given in detail in Appendix \ref{app-promt-des} and \ref{app-promt-temp}.}

\subsection{Constructing the Preliminary Test Suite}  \label{framework-B}
After obtaining the templates $\mathcal{D} = \{D_1, D_2, \dots, D_n\}$, we proceed with template instantiation, preliminary expansion of the test suite, and label verification.

At first, we utilize the tool CheckList to instantiate the templates, mapping each template $D_i$ to a set of original test cases $\mathcal{L}(D_i)$. CheckList can automatically populates templates based on predefined rules and constraints. This process generates a substantial volume of test cases, forming a raw test suite.

Then, we incorporate word-filling models for expansion to further augment the diversity and complexity of the test suite.
Specifically, we augment this set using word-filling models, replacing selected positions with \verb|[mask]| tokens
and then expanding these masked positions to broaden the scope. In this way, an expanded test case set $\mathcal{E}(\mathcal{L}(D_i))$ is obtained. The original test case set $\mathcal{L}(D_i)$ and the extended test case set $\mathcal{E}(\mathcal{L}(D_i))$ constitute an initial test suite $\mathcal{T}_o = \bigcup_{i=1}^{n} \left (\mathcal{L}(D_i) \cup \mathcal{E}(\mathcal{L}(D_i)) \right )$.

A label verification mechanism based on differential testing is designed to ensure the quality of test cases generated through the aforementioned steps.
This mechanism leverages the fundamental principle of comparing outputs from multiple models given identical inputs, thereby effectively identifying potential discrepancies and enhancing the overall reliability of the test cases.
Specifically, we implement differential testing using $N$ models $\mathcal{M} = {M_1, M_2, \dots, M_N}$. For each model $M_i$ and test case $x \in \mathcal{T}_o$, we define the model's voting function:
\begin{equation}
	\text{Vote}(M_i, x) = \mathbbm{1} \left [M_i(x) = l_{\text{ori}}(x) \right]
\end{equation}
where $M_i(x)$ denotes the predicted label of $x$ by the $i$-th model,  $l_{\text{ori}}(x)$ denotes the original label of $x$, and $\mathbbm{1}(\cdot)$ denotes the indicator function\footnote{which equals $0$ if the arguments are identical or $1$ if they differ}. Then, we calculate the consistency score of the test case $x$ as follows.
\begin{equation}
	\text{ConsistencyScore}(x) = \frac{1}{N}\sum_{i=1}^N \text{Vote}(M_i, x)
\end{equation}
Based on the consistency score, we define the label verification function $\mathcal{V}$ as:
\begin{equation}
	\mathcal{V}(x) = \begin{cases}
		0, & \text{if } \text{ConsistencyScore}(x) = 1 \\
		x, & \text{if } \text{ConsistencyScore}(x) \in (0.5, 1) \\
		x', & \text{if } \text{ConsistencyScore}(x) \leq 0.5
	\end{cases}
\end{equation}
where $x'={LLM}(x)$ denotes the optimized test case obtained by optimizing $x$ using LLMs. To be specific, for each test case $x$, we calculate its consistency score $\mathcal{V}(x)$. If $\text{ConsistencyScore}(x) = 1$, then we elect to drop the test case, as it is likely short of requisite complexity or challenge. If $\text{ConsistencyScore}(x) \in (0.5, 1)$, then $x$ is added to the test suite, as this indicates an adequate level of discriminative power and analytical value. If $\text{ConsistencyScore}(x) \leq 0.5$, then $x$ is optimized using LLMs to obtain the optimized test case $x'=\mathcal{R}(x)$, and added to the test suite.
Following the aforementioned procedures, we obtain an preliminary test suite $\mathcal{T}_1$.

\subsection{Multi-capability Expansion}  \label{sec:framework-C}

{Based on the evaluation dimensions proposed in ~ \cite{tan-etal-2021-reliability}, we redefine the dimensions of linguistic phenomena, sensitive attributes, and adversarial attacks, mapping them to taxonomy, fairness, and robustness, respectively. This approach involves enhancing each test case across multiple dimensions to increase the comprehensiveness of the evaluation.}

The taxonomy expansion emphasizes the refinement of lexical elements within sentences. Specifically, a method of selective word replacement has been employed, targeting nouns, verbs, adverbs, and adjectives within each sentence. These words are substituted with semantically related terms, including synonyms, hypernyms, and hyponyms. This serves to assess the model's capacity for understanding diverse vocabularies and interpreting varied linguistic expressions.
{In this capability expansion phase, it can lead to excessive use of related word substitutions that alter the underlying facts. For example, "{\textit{This is a delicious food.}}" may be modified to "{\textit{This is an acceptable food.}}". We will analyze mitigation strategies in Section~\ref{sec:exp-setting-basic} to preserve semantic integrity during expansion operations.}

The fairness expansion involves the insertion of diverse attributes following the subject of each sentence. These attributes encompass a wide range of demographic factors, including but not limited to place of origin, religious or philosophical beliefs, and sexual orientation. Through this expansion, it can be assessed whether the model exhibits bias when processing sentences involving different groups.

The robustness expansion can be divided into two distinct levels: preliminary robustness and adversarial robustness.
In the preliminary level, common textual variations are simulated through targeted modifications in spelling and other linguistic elements to assess the model's adaptability to minor textual perturbations.
The more advanced tier of adversarial robustness employs adversarial attack methodologies to generate challenging test cases, potentially inducing misclassifications or erroneous outputs from the model.
By implementing this multi-tier robustness testing framework, a comprehensive evaluation of the model's performance and stability when faced with both imperfect and potentially malicious inputs can be conducted.

To be specific, test suite generation takes the following procedure. After applying the taxonomy expansion ($\mathcal{T}_{tax}$), fairness expansion ($\mathcal{T}_{fair}$), and preliminary robustness expansion ($\mathcal{T}_{pre-rob}$), we merge the three test suites to obtain a comprehensive expanded test suite $\mathcal{T}_{c} = \mathcal{T}_{tax} \cup \mathcal{T}_{fair} \cup \mathcal{T}_{pre-rob}$. After that, we perform adversarial robustness extension on $\mathcal{T}_{c}$ to obtain the adversarial robustness test suite $\mathcal{T}_{adv-rob}$.

\subsection{Constructing the Final Test Suite}  \label{sec:framework-D}

\begin{figure}[h]
	\centering
	\includegraphics[width=\linewidth]{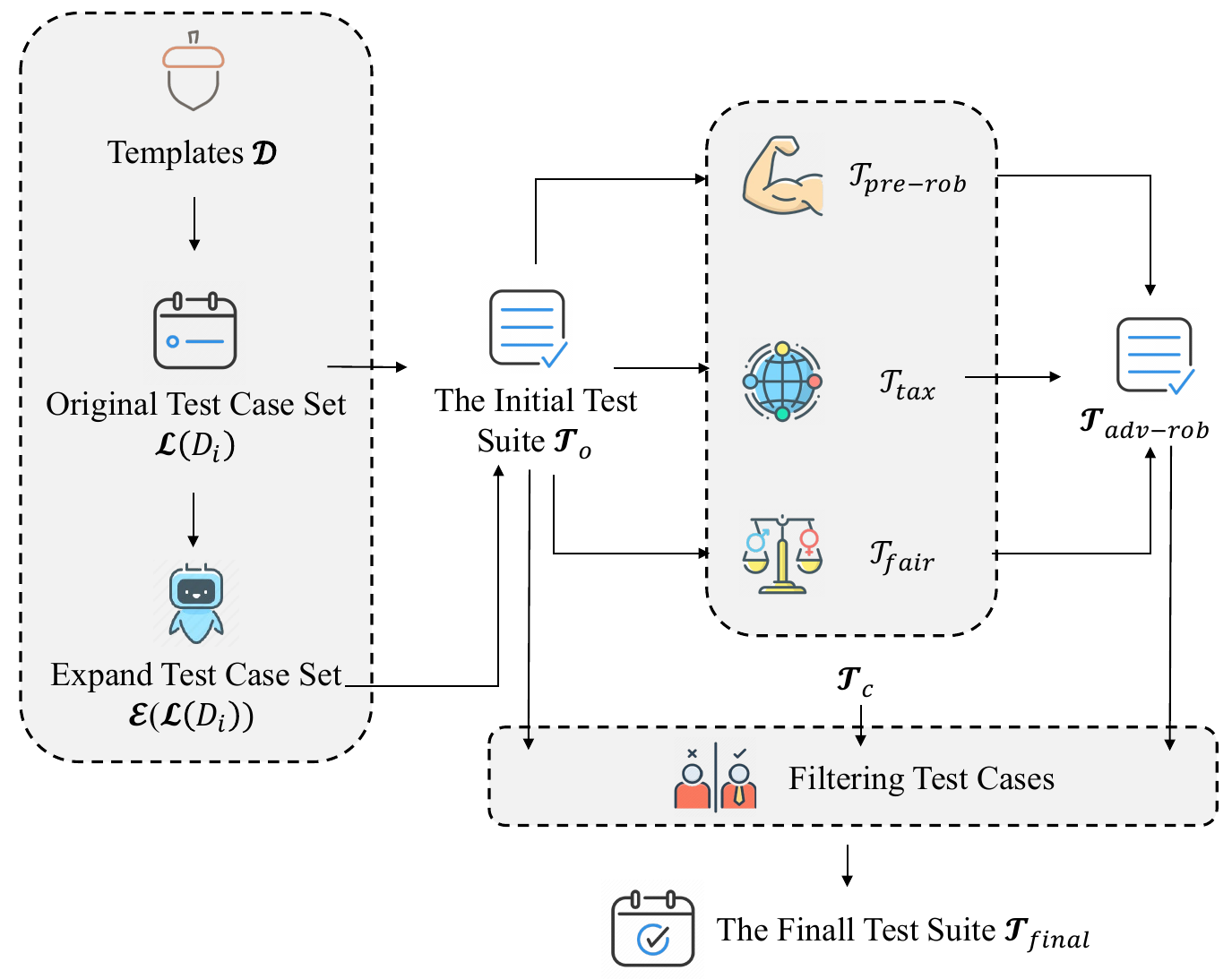}
	\caption{Overview of constructing the final test suite}
	\label{fig:constructing}
\end{figure}

In the final phase, we conduct a comprehensive integration of all previously generated test cases.
The specific process is shown in Figure \ref{fig:constructing}. In the ``filtering test cases'' part, we employ the differential testing method outlined in Section~\ref{framework-B}, with modifications to optimize test scale and execution time. Specifically,
if $ \text{ConsistencyScore}(x) < 1$, then $x$ is added to the test suite $\mathcal{T}_{final}$.
This selective process not only maintains the comprehensiveness of the evaluation, but also improves the efficiency and pertinence of the entire evaluation process.

\section{Experiments} \label{sec:exp}
We have implemented the AutoTestForge framework as an evaluation tool for NLP models. To validate its performance in real-world application scenarios, we carry out experiments on two tasks: sentiment analysis (SA) and semantic text similarity (STS).

\subsection{Research Questions} \label{sec-question}
The aim of the experiments is to answer the following research questions.

\begin{itemize}
	\item \textbf{RQ1}: How is the performance of AutoTestForge compare to existing testing approaches and methodologies?

	\item \textbf{RQ2}: What are the differences in effectiveness between human-designed templates and LLM-generated ones?
	
	\item \textbf{RQ3}: Which mainstream LLMs demonstrate superior capabilities in template generation?
	
	\item \textbf{RQ4}: How is the quality of test cases generated by LLMs?
	
	\item \textbf{RQ5}: Are the test cases generated by AutoTestForge a threat to LLMs?
\end{itemize}

Specifically,
RQ1 evaluates the effectiveness of AutoTestForge by benchmarking it against established testing approaches and methodologies in the field.
RQ2 investigates the comparative strengths and limitations between manually crafted templates and those automatically generated by LLMs.
RQ3 evaluates the template generation capabilities across prominent LLMs to determine which models excel in producing high-quality testing templates.
RQ4 performs a detailed assessment of LLM-generated test cases to verify their quality standards.
RQ5 examines the effectiveness of AutoTestForge's LLM-generated test cases in evaluating both the originating LLM and other LLMs.

\subsection{Datasets and Tools}

In order to ensure the scientific and representative evaluation of the experiment, this study selects a standard dataset that is widely recognized in related fields as the evaluation benchmark. Specifically, we use four influential datasets for sentiment analysis (SA) task as follows, covering a diverse corpus such as film reviews and business reviews.

\begin{itemize}
	\item \textbf{SST-2} (Stanford Sentiment Treebank 2)  \cite{socher-etal-2013-recursive} is a sentiment analysis benchmark dataset released by Stanford University, which focuses on binary classification tasks. It comprises approximately $70,000$ short sentences and phrases from movie reviews, with positive or negative labels.
	
	\item \textbf{Yelp Polarity}  \cite{NIPS2015_250cf8b5} is a large-scale binary classification dataset extracted from the Yelp business review dataset, containing $560,000$ training samples and $38,000$ test samples. In this dataset, $1$-star and $2$-star reviews are categorized as negative, while $4$-star and $5$-star reviews are classified as positive.
	
	\item \textbf{Rotten Tomatoes dataset}  \cite{pang-lee-2005-seeing} originates from the renowned movie review aggregation website, encompassing approximately $10,000$ short review excerpts from professional film critics. Each sample is labeled as either positive or negative.
	
	\item \textbf{IMDB}  \cite{maas-EtAl:2011:ACL-HLT2011} is a widely-used sentiment analysis dataset derived from the Internet Movie Database, containing 50,000 movie reviews evenly split into training and test sets. Each sample is labeled as either positive or negative.
\end{itemize}
For semantic textual similarity (STS) task, we use three benchmark datasets as follows, which encompass diverse textual content including question-answer pairs and news headlines.

\begin{itemize}
	\item \textbf{QQP} (Quora Question Pairs)  \cite{wang-etal-2018-glue} is a semantic similarity dataset consisting of over $400,000$ question pairs from Quora. The task involves determining whether two questions are semantically equivalent, which forms a binary classification problem to identify duplicate questions.
	
	\item \textbf{STS-B} (Semantic Textual Similarity Benchmark)  \cite{cer-etal-2017-semeval} is a collection of sentence pairs drawn from various sources including news headlines, video captions, and image descriptions. It contains approximately $8,600$ sentence pairs scored on a scale from $0$ to $5$, indicating their semantic similarity. In our experiments, we consider data with a value less than $2.5$ as dissimilar and data with a value greater than or equal to $2.5$ as similar.
	
	\item \textbf{MRPC} (Microsoft Research Paraphrase Corpus)  \cite{dolan-brockett-2005-automatically} is a dataset compiled from news sources, comprising $5,801$ sentence pairs annotated by human experts. Each pair is labeled to indicate whether the sentences are semantically equivalent, serving as a benchmark for paraphrase identification tasks.
\end{itemize}

To ensure the consistency in sample size across the datasets, $1,820$ samples are uniformly selected from each dataset, matching the number of samples in the SST-2 test set.
Additionally, our approach is evaluated against two state-of-the-art tools in behavioral testing as follows.

\begin{itemize}
	\item \textbf{CheckList} \cite{ribeiroAccuracyBehavioraltesting2020} has been demonstrated a versatile tool applicable to a wide range of NLP tasks. It provides a more comprehensive evaluation of model performance across diverse scenarios. Through its application to multiple tasks, CheckList has been shown effective in identifying critical flaws in both commercial and NLP models.
	
	\item \textbf{TestAug} \cite{yangTestAugFrameworkAugmenting2022} leverages the GPT-3 engine for test case generation in behavioral testing of NLP models. A classifier is employed by TestAug to filter out invalid GPT-3 outputs, and the valid ones are expanded into templates for generating additional test cases. The advantage of TestAug over other methods in terms of bug discovery, test diversity, and efficiency has been demonstrated through experimental results.
\end{itemize}

\subsection{Experimental Setting}
\subsubsection{Basic setting} \label{sec:exp-setting-basic}

\begin{table*}[htbp]
	\centering
	\caption{Details of the adversarial attack methods. "Level" indicates attack granularity, while "Goal Function" defines the attack objective. "Constraints-Enforced" specifies the limitations that maintain text boundaries and coherence. "Transformation" details permissible text modifications for creating adversarial examples, and "Search Method" describes the approach for identifying successful transformations. }
	\label{tab-textattack}
	\resizebox{\textwidth}{!}{
		\begin{tabular}{cccccc}
			\toprule
			Attack Recipe Name & Level                       & Goal Function                                                                      & Constraints-Enforced                                                               & Transformation                                                                                                                                  & Search Method                                                          \\ \hline
			Deepwordbug \cite{JiDeepWordBug18}        & character-level             & \begin{tabular}[c]{@{}c@{}}\{Untargeted, Targeted\} \\ Classification\end{tabular} & \begin{tabular}[c]{@{}c@{}}Levenshtein edit\\ distance\end{tabular}                & \begin{tabular}[c]{@{}l@{}}\{Character Insertion, \\ Character Deletion, \\ Neighboring Character Swap,\\ Character Substitution\}\end{tabular} & Greedy-WIR                                                             \\ \hline
			PSO \cite{zang-etal-2020-word}               & word-level                  & \begin{tabular}[c]{@{}c@{}}Untargeted \\ Classification\end{tabular}               &     /                                                                               & HowNet Word Swap                                                                                                                                  & \begin{tabular}[c]{@{}c@{}}Particle Swarm \\ Optimization\end{tabular} \\ \hline
			Textbugger \cite{ndss-LiJDLW19}        & character \& word-level 	 & \begin{tabular}[c]{@{}c@{}}Untargeted \\ Classification\end{tabular}               & \begin{tabular}[c]{@{}c@{}}USE sentence encoding \\ cosine similarity\end{tabular} & \begin{tabular}[c]{@{}l@{}}\{Character Insertion, \\ Character Deletion,\\ Neighboring Character Swap, \\ Character Substitution\}\end{tabular} & Greedy-WIR                             \\ \bottomrule
		\end{tabular}
	}
\end{table*}

{We evaluate AutoTestForge on sentiment analysis using three leading LLMs to generate templates: GPT-4\footnote{\url{https://openai.com/index/gpt-4/}}, Llama3-70B\footnote{\url{https://llama.meta.com/llama3/}}, and Claude3-Opus\footnote{\url{ https://www.anthropic.com/news/claude-3-family}}. Additionally, we set the random seed to $42$ for experimental reproducibility. }

To address the sentiment and semantic integrity challenges discussed in Section \ref{sec:framework-C} during taxonomy expansion, we propose a solution that utilizes roberta-base for scoring. After identifying candidate words, we leverage roberta-base to compute a score for each word. The selection process then filters for candidate words where the score difference from the current word remains below a threshold of $1$.
At the same time, for hyponyms, we only search for words with a depth of less than $3$. These settings alleviate this concern to a certain extent.
And AutoTestForge implements fairness considerations including kin color, sexual orientation, religious beliefs, occupations, and nationalities in the fairness capability expansion phase.

To optimize large-scale test case generation, we implement the following strategy. Each template generates $500$ samples, with $20\%$ randomly selected. For each selected sample, we randomly choose $5$ words for mask replacement, creating new templates. Each new mask template subsequently generates $10$ samples. We employ the roberta-large model for word-filling. The model has superior performance in contextual understanding and appropriate vocabulary generation.

\begin{table}[htbp]
	\centering
	\caption{Models for differential testing. Everyone can use the Checkpoint Identifier to find the corresponding model on the Huggingface.}
	\label{tab-diff-model}
	\resizebox{\linewidth}{!}{
		\begin{tabular}{cccc}
			\toprule
			Model name (Checkpoint Identifier)			& Task  	&  Training set		& Accuracy 		 	\\ \hline
			textattack/bert-base-uncased-SST-2		  	& SA 		& SST-2           	& $93.98\%$   	 	\\
			textattack/bert-base-uncased-yelp-polarity	& SA 		& Yelp-polarity   	& $96.99\%$ 		\\
			textattack/roberta-base-SST-2      			& SA		& SST-2           	& $95.44\% $        \\
			textattack/roberta-base-rotten-tomatoes		& SA		& Rotten-tomatoes 	& $90.34\% $        \\
			textattack/xlnet-base-cased-imdb        	& SA        & imdb            	& $95.35\%$ 		\\
			
			Intel/bert-base-uncased-mrpc 				& STS		& MRPC           	& $86.03\%$    		\\
			textattack/bert-base-uncased-STS-B			& STS		& STS-B   			& $86.00\%$ 		\\
			textattack/roberta-base-MRPC      			& STS		& MRPC           	& $91.18\% $        \\
			JeremiahZ/roberta-base-qqp					& STS		& QQP 				& $91.53\% $        \\
			textattack/xlnet-base-cased-QQP             & STS		& QQP           	& $90.66\%$ \\ \bottomrule
		\end{tabular}
	}
	
\end{table}

We employ five differential testing models in different task, encompassing diverse NLP architectures and widely-used test sets. Table \ref{tab-diff-model} lists the specific performance metrics of these models.
Specifically, the columns Model name, Training set, and Accuracy show the architecture of the model, the datasets on which each model is trained, and the accuracy of the model on the corresponding dataset, respectively.
For samples that produce output labels different from the original labels, GPT-3.5-TURBO\footnote{\url{https://openai.com/chatgpt}} is adopted for refinement.

We consider three adversarial attack methods: character-level (DeepWordBug \cite{JiDeepWordBug18}), word-level (PSO-based Adversarial Example Search Algorithm \cite{zang-etal-2020-word}), and combined character \& word-level (Textbugger \cite{ndss-LiJDLW19}) attacks.
Table \ref{tab-textattack} provides a comprehensive overview of these methods.  
Additional methods conforming to TextAttack~\cite{morrisTextAttackframeworkadversarial2020a} specifications can be readily modified and incorporated into the framework.

\subsubsection{Advanced setting}
To comprehensively assess AutoTestForge's potential and compare the capabilities of LLMs and humans in designing error detection templates, we establish three evaluation modes to assess template generation capabilities:
the Human \& Human mode means human experts creating both descriptions and corresponding templates;
the Human \& AI mode means human experts develop the descriptions while LLMs generate the associated templates;
and the AI \& AI mode means LLMs produce both the descriptions and their respective templates.

To ensure the scientific rigor and comparability of the experiments, we adopt the descriptions and templates provided in TestAug as the benchmark for human expert design. \review{However, during the implementation, we discover instances of label inconsistencies in TestAug's test cases. Therefore, we employ the differential testing method discussed in Section \ref{framework-B} to filter these cases. Following this validation process, we establish a consistent structure where each label is associated with $6$ sentence structure descriptions, each corresponding to $3$ templates.}

\subsubsection{Models for Evaluation}
To comprehensively evaluate the effectiveness of the framework, we consider various aspects including model architecture, parameter scale, and case sensitivity.
Specifically, we select four models for each of the two tasks. The sentiment analysis models are trained on the SST dataset, while the semantic textual similarity Task models are trained on the QQP dataset. Detailed specifications for these models are presented in Table \ref{tab-eva-model}. Additionally, to address Research Question $5$, we incorporate four large language models into our evaluation: GPT-3.5-TURBO, GPT-4o, Llama3-8B and Llama3-70B. The prompt for performing two tasks on  LLMs is given in Appendix \ref{app-prrompt-q5}.

\begin{table}[h]
\centering
\caption{Models for Evaluation.}
\label{tab-eva-model}
\resizebox{\linewidth}{!}{
	\begin{tabular}{ccc}
		\toprule
		Model name                  & Task & Checkpoint Identifier                                               \\ \hline
		Distilbert-base-cased-SST-2 & SA   & textattack/distilbert-base-cased-SST-2 \\
		Albert-base-v2-SST-2        & SA   & textattack/albert-base-v2-SST-2 \\
		Bert-base-uncased-SST-2     & SA   & textattack/bert-base-uncased-SST-2    \\
		Roberta-base-SST-2          & SA   & textattack/roberta-base-SST-2            \\
		Albert-base-v2-QQP          & STS  & textattack/albert-base-v2-QQP                      \\
		Bert-base-uncased-QQP       & STS  & textattack/bert-base-uncased-QQP   \\
		Roberta-base-qqp            & STS  & JeremiahZ/roberta-base-qqp                       \\
		Xlnet-base-cased-QQP        & STS  & textattack/xlnet-base-cased-QQP           \\
		\bottomrule
	\end{tabular}
}
\end{table}

\subsection{Results Analysis} 
{This section empirically addresses the research questions raised at the beginning of this section.}
We systematically analyze each question, presenting corresponding experimental data and observations to substantiate the conclusions.

\subsubsection{Performance comparison with existing methods}

\begin{table*}[htbp]
	\caption{\review{The comparison of the bug detection capabilities between datasets, tools, and AutoTestForge. Each cell shows the failure rate, where a higher percentage indicates better error detection capability.} }
	\centering
	\label{tab1}
	\resizebox{\textwidth}{!}{
		\begin{tabular}{c|ccc|c|cc|c|ccc|c}
			\toprule
			\multicolumn{12}{c}{Sentiment Analysis Task (SA)}                   \\  \hline
			
			\multirow{2}{*}{Model Type} & \multicolumn{4}{c|}{DataSet}                        & \multicolumn{3}{c|}{Tool}      & \multicolumn{4}{c}{AutoTestForge}                 \\  \cline{2-12}
			& SST-2  & Yelp Polarity & Rotten Tomatoes & \textbf{Avg}     & CheckList & TestAug & \textbf{Avg}     & Human \& Human & Human \& AI & AI \& AI & \textbf{Avg}     \\  		\hline
			Distilbert-base-cased-SST-2 & $8.02\%$ & $11.87\%$       & $10.41\%$         & $10.10\%$ & $28.72\%$   & $23.21\%$ & $25.96\%$ & $33.91\%$        & $30.02\%$     & $35.34\%$  & $33.09\%$ \\
			Albert-base-v2-SST-2        & $8.68\%$ & $11.92\%$       & $10.98\%$         & $10.52\%$ & $17.02\%$   & $21.44\%$ & $19.23\% $& $34.17\%$        & $26.10\% $    & $29.21\%$  & $29.83\%$ \\
			Bert-base-uncased-SST-2     & $6.70\%$ & $8.63\%$        & $10.13\% $        & $8.49\%$  & $13.29\%$   & $15.47\%$ & $14.38\%$ & $31.08\%$        & $35.74\%$     & $37.62\%$  & $34.81\%$ \\
			Roberta-base-SST-2          & $4.56\%$ & $5.05\%$        & $9.57\%$          & $6.39\%$  & $4.83\%$    & $10.90\%$ & $7.87\% $ & $29.30\%$        & $24.27\%$     & $23.87\%$  & $25.81\%$ \\  \hline
			Avg                   & $6.99\%$ & $9.37\%$        & $10.27\%$         & $8.88\%$  & $15.97\%$   & $17.75\% $& $16.86\%$ & $32.11\%$        & $29.03\%$     & $31.51\%$  & $30.89\%$ \\   \hline
			\multicolumn{12}{c}{Semantic Textual Similarity Task (STS)}                         \\ \hline
			\multirow{2}{*}{Model Type}                          & \multicolumn{4}{c|}{DataSet}   & \multicolumn{3}{c|}{Tool}     & \multicolumn{4}{c}{AutoTestForge}                                                                                             \\ \cline{2-12}
			& QQP    & MRPC    & STS-B & \textbf{Avg}     & CheckList & TestAug & \textbf{Avg}     & Human \& Human & Human \& AI & AI \& AI & \textbf{Avg}     \\   \hline
			Albert-base-v2-QQP     & $9.53\%$ & $36.86\%$ & $25.33\%$   & $23.91\%$  & $32.58\%  $ & $24.69\%$   & $28.63\%$   & $35.39\%$  & $39.94\%$   & $38.57\%$  & $37.97\% $                \\
			Bert-base-uncased-QQP  & $9.38\%$ & $37.68\%$ & $24.29\%$   & $23.78\%$  & $25.86\% $  & $25.70\%  $ & $25.78\%$   & $39.23\%$  & $35.95\%$   & $37.53\%$  & $37.57\%$                 \\
			Roberta-base-QQP 	   & $8.44\%$ & $26.91\%$ & $25.00\%$   & $20.12\%$  & $8.75\% $   & $7.97\%   $ & $8.36\%$    & $33.66\%$  & $27.60\%$   & $30.32\%$  & $30.53\%$                 \\
			Xlnet-base-cased-QQP   & $9.14\%$ & $39.90\%$ & $25.05\%$   & $24.70\%$  & $4.14\%$    & $5.23\%  $  & $4.69\%$    & $39.01\%$  & $28.03\%$   & $29.76\%$  & $32.27\%$                 \\ \hline
			Avg                    & $9.12\%$ & $35.34\%$ & $24.92\%$   & $23.13\%$  & $17.83\%$   & $15.90\%$   & $16.87\%$   & $36.82\%$  & $32.88\%$   & $34.04\%$  & $34.58\%$                \\ \bottomrule
			
		\end{tabular}
	}
\end{table*}

We conduct systematic comparative experiments to evaluate AutoTestForge against testing tools and widely adopted datasets, assessing its error detection capabilities across multiple models. The detailed results are presented in Table \ref{tab1}. It can be observed that AutoTestForge exhibits performance advantages in both sentiment analysis and semantic textual similarity tasks.

Specifically, AutoTestForge achieves average error detection rates that surpass current popular datasets by $22.01\%$ ($30.89\%$ vs. $8.88\%$) across both tasks. When compared to advanced behavioral testing tools, AutoTestForge demonstrates equally detection capabilities. Relative to CheckList, AutoTestForge show improvements of $14.92\%$ ($30.89\%$ vs. $15.97\%$) and $16.75\%$ ($34.58\%$ vs. $17.83\%$) across the two tasks. Similarly, when compared to TestAug, it exhibits improvements of $13.14\%$ ($30.89\%$ vs. $17.75\%$) and $18.66\%$ ($34.58\%$ vs. $15.90\%$).

Experimental results show that the three test case generation modes of AutoTestForge perform effectively across different model architectures.
For the sentiment analysis task, the AI \& AI mode achieves a peak error detection rate of $37.62\%$ on the Bert-base-uncased-SST-2 model. For the semantic textual similarity task, the Human \& AI mode reaches a detection rate of $39.94\%$ on the Albert-base-v2-QQP model, demonstrating remarkable testing effectiveness.
For both types of tasks, AutoTestForge's Human \& Human and AI \& AI modes exhibit comparable performance, with a maximum difference of only $2.28\%$. These results not only validate the method's reliability but also indicate that both manually designed and AI-assisted generated test templates can effectively identify model defects. Particularly in complex tasks such as semantic similarity, AutoTestForge's error detection capabilities outperform traditional methods more evidently, highlighting its practical value in advanced NLP model testing.

\begin{tcolorbox}[colback=gray!10,colframe=black,arc=2mm, auto outer arc,title={Answer to RQ1},breakable]
	AutoTestForge outperforms existing approaches, achieving error detection rates that surpass traditional datasets and showing improvements over state-of-the-art testing tools like CheckList and TestAug across both sentiment analysis and semantic textual similarity tasks.
\end{tcolorbox}

\subsubsection{Evaluating LLM-generated and human-designed templates} \label{sec-eva-template}

\begin{table}[htbp]
	\caption{Comparison in error detection rates: only using CheckList for word completion versus using AutoTestForge to expand the test suite.}
	\centering
	\label{tab2}
	\resizebox{\linewidth}{!}{
		\begin{tabular}{clccccc}
			\toprule
			\multicolumn{7}{c}{Sentiment Analysis Task (SA)}               \\ \hline			
			\multicolumn{2}{c}{Data Source}                        & Distilbert & Albert   & Bert & Roberta     & Avg     \\ \hline
			\multirow{4}{*}{CheckList} & Human \& Human 			& $28.72\%  $             & $17.02\%$          & $13.29\%$           & $4.83\%$           & $15.97\%$ \\
			& Human \& AI         									& $34.16\%$               & $14.17\% $         & $15.27\%$           & $24.66\% $         & $22.06\%$ \\
			& AI \& AI             									& $30.02\%$               & $26.10\%$          & $35.74\%$           & $24.27\%$          & $29.03\%$ \\ \cline{2-7}
			& Avg                									& $30.96\% $              & $19.10\%$          & $21.43\%$           & $17.92\%$          & $22.35\%$ \\ \hline
			\multirow{4}{*}{AutoTestForge}   & Human \& Human     & $33.91\%$               & \textbf{$34.17\%$} & $31.08\%$           & \textbf{$29.30\%$} & $32.11\%$ \\
			& Human \& AI         									& $30.02\% $              & $26.10\%$          & $35.74\%$           & $37.605\%$          & $29.03\%$ \\
			& AI \& AI             									& \textbf{$35.34\%$}      & $29.21\% $         & \textbf{$37.62\%$}  & $23.87\%$          & $31.51\%$ \\ \cline{2-7}
			& Avg                									& $34.58\%$               & $27.90\%$          & $34.82\% $          & $24.81\% $         & $30.53\% $ \\ \hline
			\multicolumn{7}{c}{Semantic TextTual Similarity Task (STS)}                                       \\ \hline
			\multicolumn{2}{c}{Data Source}                        & Albert  & Bert    & Roberta & Xlnet   & Avg     \\ \hline
			\multirow{4}{*}{CheckList}     & Human \& Human  & $ 32.58\% $ & $ 25.86\% $ & $ 8.75\%  $ & $ 4.14\%  $ & $ 17.83\%$ \\
			& Human \& AI     & $ 36.59\% $ & $ 34.85\% $ & $ 14.69\% $ & $ 13.86\% $ & $ 25.00\%$ \\
			& AI \& AI        & $ 35.78\% $ & $ 44.14\% $ & $ 12.27\% $ & $ 14.38\% $ & $ 26.64\%$ \\ \cline{2-7}
			& Avg             & $ 34.98\% $ & $ 34.95\% $ & $ 11.90\% $ & $ 10.79\% $ & $ 23.16\%$ \\ \hline
			\multirow{4}{*}{AutoTestForge} & Human \& Human  & $ 35.39\% $ & $ 39.23\% $ & $ 33.66\% $ & $ 39.01\% $ & $ 36.82\%$ \\
			& Human \& AI     & $ 39.94\% $ & $ 35.95\% $ & $ 27.60\% $ & $ 28.03\% $ & $ 32.88\%$ \\
			& AI \& AI        & $ 38.57\% $ & $ 37.53\% $ & $ 30.32\% $ & $ 29.76\% $ & $ 34.04\%$ \\ \cline{2-7}
			& Avg             & $ 37.97\% $ & $ 37.57\% $ & $ 30.53\% $ & $ 32.27\% $ & $ 34.58\%$ \\
			\bottomrule
		\end{tabular}
}
\end{table}

\begin{table*}[htbp]
	\caption{	
		\review{Comparison of error detection rates for test templates generated by different Large Language Models (GPT-4, Claude-3-Opus, and Llama3-70B) under different generation strategies (Human \& AI and AI \& AI). The "CheckList" column represents test cases derived through word substitution and label filtering of LLM-generated templates, whereas the "AutoTestForge" column represents test cases obtained through the workflow illustrated in Figure \ref{fig:AutoTestForge}.}
	}
	\centering
	\label{tab3}
	\resizebox{\textwidth}{!}{
		
		\begin{tabular}{c|cccc|cccc|cccc}
			\toprule
			\multirow{3}{*}{ Model Type} & \multicolumn{4}{c|}{GPT-4}                                   & \multicolumn{4}{c|}{Claude-3-Opus}                            & \multicolumn{4}{c}{Llama3-70B}                              \\ \cline{2-13}
			& \multicolumn{2}{c}{Human\&AI} & \multicolumn{2}{c|}{AI\&AI} & \multicolumn{2}{c}{Human\&AI} & \multicolumn{2}{c|}{AI\&AI} & \multicolumn{2}{c}{Human\&AI} & \multicolumn{2}{c}{AI\&AI} \\  \cline{2-13}
			& checklist       & AutoTestForge      & checklist     & AutoTestForge    & checklist       & AutoTestForge      & checklist     & AutoTestForge    & checklist       & AutoTestForge      & checklist     & AutoTestForge    \\   \hline
			\multicolumn{13}{c}{Sentiment Analysis Task (SA)}                                                                                                                                                                  \\ \hline
			Distilbert-base-cased-SST2 &$ 12.85\%$         & $30.02\%$      & $20.78\%$       & $35.34\%$    & $34.16\%$         & $27.18\%$      & $16.14\%$       & $30.04\%$    & $18.23\%$         & $27.26\%$      & $34.49\%$       & $23.06\%$    \\
			Albert-base-v2-SST2        & $13.95\%  $       & $26.10\%$      & $15.30\%$       & $29.21\%$    & $14.17\%$         & $19.06\%$      & $6.75\% $       & $15.21\% $   & $12.25\%$         & $21.04\%$      & $20.32\%$       & $21.86\%$    \\
			Bert-base-uncased-SST2     & $15.10\% $        & $35.74\%$      & $18.59\%$       & $31.36\% $   & $15.27\%$         & $30.37\%$      & $21.25\%$       & $37.62\%$    & $12.03\% $        & $30.51\%$      & $35.75\%$       & $26.58\% $   \\
			Roberta-base-SST2          &$ 10.49\%$         & $21.71\%$      & $6.41\%$        &$ 23.87\%$    & $24.66\%$         & $24.27\%$      & $2.42\%$        & $13.45\%$    & $6.15\%$          & $20.77\%$      & $21.25\% $      & $20.59\%$    \\ \hline
			Avg                   & $13.10\%$         & $28.39\%$      & $15.27\%$       & $29.95\%$    & $22.06\%$         & $25.22\% $     & $11.64\% $      & $24.08\%$    & $12.16\% $        & $24.90\%$      & $27.95\%$       & $23.02\%$    \\  \hline
			\multicolumn{13}{c}{Semantic Textual Similarity Task (STS)}     \\   \hline
			Albert-base-v2-QQP    & $36.59\%$ & $29.27\%$ & $35.78\%$ & $34.81\%$ & $20.77\%$ & $27.71\%$ & $22.15\%$ & $32.72\%$ & $29.45\%$ & $39.94\%$ & $25.87\%$ & $38.57\%$ \\
			Bert-base-uncased-QQP & $34.85\%$ & $33.33\%$ & $44.14\%$ & $37.53\%$ & $27.59\%$ & $28.99\%$ & $23.07\%$ & $30.97\%$ & $31.17\%$ & $35.95\%$ & $16.21\%$ & $32.11\%$ \\
			Roberta-base-QQP      & $12.88\%$ & $23.75\%$ & $12.27\%$ & $26.26\%$ & $10.11\%$ & $21.15\%$ & $9.50\% $ & $22.50\%$ & $14.69\%$ & $27.60\%$ & $9.58\% $ & $30.32\%$ \\
			Xlnet-base-cased-QQP  & $13.86\%$ & $26.43\%$ & $14.38\%$ & $29.76\%$ & $6.67\%$  & $20.16\%$ & $0.69\% $ & $17.75\%$ & $10.94\%$ & $28.03\%$ & $8.97\% $ & $26.67\%$ \\ \hline
			Avg                   & $24.55\%$ & $28.20\%$ & $26.64\%$ & $32.09\%$ & $16.28\%$ & $24.50\%$ & $13.85\%$ & $25.99\%$ & $21.56\%$ & $32.88\%$ & $15.16\%$ & $31.92\%$ \\ \bottomrule
		\end{tabular}
	}
\end{table*}

We provide a comprehensive comparison between the creativity of humans and LLMs, with the experimental results shown in Table \ref{tab2}. Note that CheckList directly uses the original templates for testing, while AutoTestForge conducts testing after the expansion process. We compare the performance of human experts and LLMs in different scenarios.

The experimental results demonstrate the advantages of AI in generating test templates. Compared to pure human design, introducing AI can improve the error detection capability of templates. Specifically, for the two tasks, templates generated solely by AI achieve $6.09\% $ ($22.06\%$ vs. $15.97\%$) and $ 7.17\%$ ($25.00\%$ vs. $17.83\%$) increases in error detection rate than human-designed ones; when the entire process is AI-led, this advantage rises to $13.06\%$ ($23.03\%$ vs. $15.97\%$) and $8.75\%$ ($26.64\%$ vs. $17.83\%$).
This is because AI can automatically extract key features of templates from massive data, generating more diverse and comprehensive test templates. In fact, AI-generated templates are not only more numerous but also of higher quality, capable of revealing more potential defects in the system under test. In contrast, human-designed templates are limited by personal experience and perspective, making it difficult to comprehensively consider all possible test scenarios, resulting in relatively lower error detection rates.

In addition, it can be observed that AutoTestForge's improvement on human templates far exceeds its enhancement of AI-generated templates, with an improvement capability of more than $100\%$. This finding has important implications: human-designed test templates still have enormous optimization space to be explored, and the current design and application methods have not fully tapped their potential.
In comparison, AI-generated test templates already possess a high degree of maturity, making it difficult to achieve qualitative improvement through simple automatic expansion. Nevertheless, AutoTestForge can still uncover more fine-grained test scenarios for AI templates, further enriching the testing dimensions.

\begin{tcolorbox}[colback=gray!10,colframe=black,arc=2mm, auto outer arc,title={Answer to RQ2},breakable,]
	Experimental results show that AI has a stronger comprehensive capabilities than human experts in generating and applying test templates. Whether using native AI templates or AI templates expanded through AutoTestForge, the error detection rates are higher on average compared to purely human-designed ones.
\end{tcolorbox}

\subsubsection{Identifying optimal LLM for template generation } \label{app-results-LLM-generate-template}

\begin{table}[]
	\caption{\review{Summary of error detection rates across different tasks and generation strategies for three LLMs (Table \ref{tab3}).}}
	\centering
	\label{tab4}
	\resizebox{\linewidth}{!}{
		\begin{tabular}{cc|cc|cc|ccc}
			\toprule							
			\multicolumn{2}{c|}{\multirow{2}{*}{Data Source}} & \multicolumn{2}{c|}{Human\&AI} & \multicolumn{2}{c|}{AI\&AI} & \multicolumn{3}{c}{Avg}        \\ \cline{3-9}
			\multicolumn{2}{c|}{}                      & SA            & STS           & SA           & STS         & SA      & STS     & SA and STS \\ \hline
			\multirow{3}{*}{GPT-4}     	& Checklist     & 13.10\%       & 24.55\%       & 15.27\%      & 26.64\%     & 14.18\% & 25.60\% & 19.89\%    \\
			& AutoTestForge & 28.39\%       & 28.20\%       & 29.95\%      & 32.09\%     & 29.17\% & 30.15\% & 29.66\%    \\ \cline{2-9}
			& Avg           & 20.74\%       & 26.38\%       & 22.61\%      & 29.37\%     & 21.68\% & 27.87\% & 24.77\%    \\ \hline
			\multirow{3}{*}{Claude3-Opus}  	& Checklist     & 22.06\%       & 16.28\%       & 11.64\%      & 13.85\%     & 16.85\% & 15.07\% & 15.96\%    \\
			& AutoTestForge & 25.22\%       & 24.50\%       & 24.08\%      & 25.09\%     & 24.65\% & 24.80\% & 24.72\%    \\ \cline{2-9}
			& Avg           & 23.64\%       & 20.39\%       & 17.86\%      & 19.47\%     & 20.75\% & 19.93\% & 20.34\%    \\ \hline
			\multirow{3}{*}{Llama3-70B}   	& Checklist     & 12.16\%       & 21.56\%       & 27.95\%      & 15.16\%     & 20.06\% & 18.36\% & 19.21\%    \\
			& AutoTestForge & 24.90\%       & 32.88\%       & 23.02\%      & 31.90\%     & 23.96\% & 32.39\% & 28.17\%    \\ \cline{2-9}
			& Avg           & 18.53\%       & 27.22\%       & 25.49\%      & 23.53\%     & 22.01\% & 25.38\% & 23.69\%    \\					
			\bottomrule
		\end{tabular}
	}
\end{table}

We conduct a comparative analysis of test template generation quality across three leading large language models (GPT-4, Claude-3-Opus, and Llama3-70B). The experimental results are shown in Tables ~\ref{tab3} and \ref{tab4}.

In evaluating the overall error detection rates, the three models exhibit different performance. GPT-4 exhibits superior performance, achieving an average error detection rate of $24.77\%$ across all tasks and strategies. Llama3-70B follows closely with an average detection rate of $23.69\%$, while Claude3-Opus shows an average performance of $20.34\%$. Notably, the AutoTestForge method show advantages over the traditional CheckList approach across all the three models. For instance, GPT-4 achieves an average error detection rate of $29.66\%$ using AutoTestForge, evidently higher than that of $19.89\%$ achieved using CheckList.

When examining the test scenarios that exclusively use LLM-generated templates (CheckList), we observe a delicate phenomenon: the models show differences in performance across the two main tasks. In the SA task, Llama3-70B achieves the highest error detection rate of $27.95\%$ under the AI \& AI mode, while in the STS task, GPT-4 achieves optimal performance with a $26.64\%$ detection rate under the AI \& AI mode. These task-specific performance variations indicate that careful evaluation and screening processes are necessary when selecting appropriate LLMs for specific tasks.

The AutoTestForge method provides an effective solution to these challenges. As Table \ref{tab4} shows, the method consistently delivers performance improvements across different modes. Specifically, across all the three models, AutoTestForge maintains detection rates consistently above $23\%$, reaching as high as $32.39\%$, demonstrating stability. In contrast, the CheckList method shows greater variability, with detection rates fluctuating between $14.18\%$ and $25.60\%$. This result indicates AutoTestForge not only provides higher detection rates but also maintains this performance improvement consistently across different models and tasks.

\begin{tcolorbox}[colback=gray!10,colframe=black,arc=2mm, auto outer arc,title={Answer to RQ3},breakable]
	GPT-4 demonstrates superior template generation capabilities, achieving the highest overall error detection rate of $29.66\%$ when enhanced with AutoTestForge. At the same time, AutoTestForge maintains a certain degree of stability in different generative models and tasks.
\end{tcolorbox}

\subsubsection{Evaluating the quality of test cases}

\begin{table}[]
	\caption{\review{Quality distribution analysis of test cases in SA tasks, comparing the original datasets, Human-generated (CheckList) and LLM-generated test cases , and AutoTestForge-built test cases.}}
	\centering
	\label{tab-quality}
	\resizebox{\linewidth}{!}{
		\begin{tabular}{ccccc}
			\hline
			\multicolumn{2}{c}{Data Source}                   & \multicolumn{1}{c}{High} 	& \multicolumn{1}{c}{Medium} 	& \multicolumn{1}{c}{Low} \\ \hline
			\multirow{2}{*}{Human \& Human} & SST-2             & $7.74\%$                   & $86.38\%$                     & $5.88\%$                  \\
			& CheckList              &$ 0.00\%$                   & $81.05\%$                     & $18.95\%$                 \\ \hline
			\multirow{2}{*}{Human \& AI} & Original     & $6.64\%$                   & $91.10\%$                     & $2.25\%$                  \\
			& AutoTestForge 						   & $1.04\%$                   & $51.62\%$                     & $47.34\%$                 \\ \hline
			\multirow{2}{*}{AI \& AI}    & Original     & $29.32\%$                  & $69.96\%$                     & $0.71\%$                  \\
			& AutoTestForge 						   & $8.13\%$                   & $59.97\%$                     & $31.91\%$                \\
			\hline
		\end{tabular}
	}
\end{table}

We use NVIDIA's quality classifier\footnote{\url{https://huggingface.co/nvidia/quality-classifier-deberta}} to evaluate test cases generated by Claude3-Opus in the SA task. This classifier, built on the DeBERTa-V3 Base model, categorizes text into "high", "medium", and "low" quality levels. The model is trained on $22,828$ Common Crawl text samples and achieves $82.52\%$ accuracy on $7,128$ evaluation samples. As is shown in Table~\ref{tab-quality}, the quality assessment reveals distinct patterns across different test case scenarios.

In the human-generated test cases, most of the SST-2 dataset and CheckList-generated cases fall into the medium-quality category. Notably, SST-2 maintains $7.74\%$ high-quality cases, while CheckList produces no high-quality examples and generates a higher proportion ($18.95\%$) of low-quality cases. This may be attributed to the oversimplification of human-designed templates, resulting in a decrease in the test cases' quality.
In the Human \& AI mode, original human-designed cases are predominantly medium-quality ($91.10\%$) with few low-quality cases ($2.25\%$). However, after AutoTestForge's processing, the quality distribution shifted: low-quality cases increases to $47.34\%$ while medium-quality cases decreases to $51.62\%$.
In the AI \& AI scenario, LLM-generated original test cases exhibit the best performance, with high-quality cases reaching $29.32\%$. This confirms LLM's superior language generation capabilities. However, after applying AutoTestForge, the proportion of high-quality cases decreases to $8.13\%$.

Overall, the experimental results indicate that LLM-generated test cases generally have higher quality than human-designed ones. Besides, that AutoTestForge's robustness enhancement mechanism leads to a decline in quality metrics, which aligns with its design goal of testing model robustness through perturbations and edge cases.

\begin{tcolorbox}[colback=gray!10,colframe=black,arc=2mm, auto outer arc,title={Answer to RQ4},breakable]
	Experimental results indicate that LLM-generated test cases generally have higher quality than human-designed ones.
\end{tcolorbox}

\subsubsection{Evaluate the performance of AutoTestForge on the LLMs} \label{sec-q-5}

\begin{figure*}[h]
	\centering
	\subfigure[Llama3-70b-8192]{
		\includegraphics[width=0.3\textwidth]{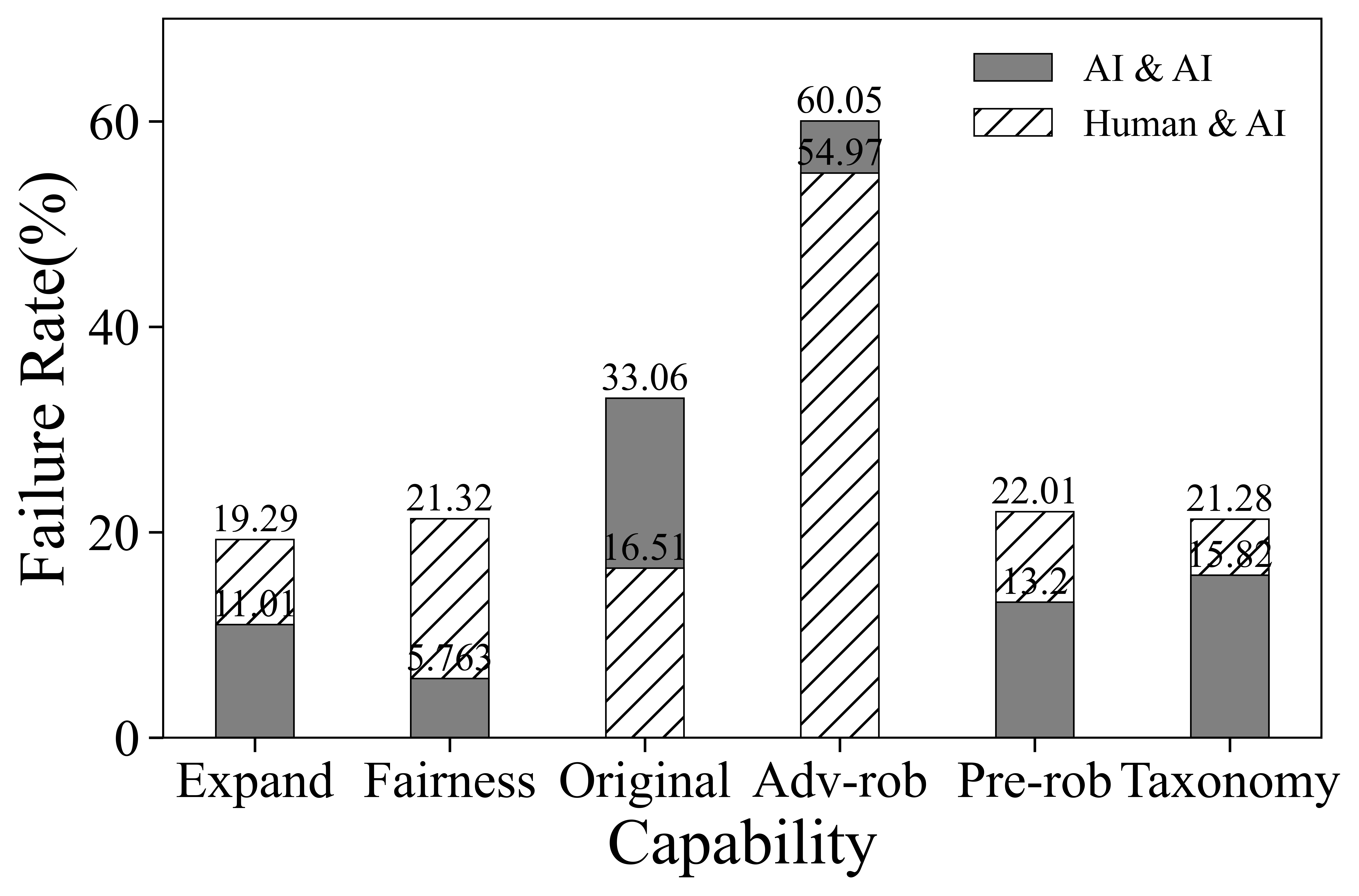}
	}
	\subfigure[GPT-4]{
		\includegraphics[width=0.3\textwidth]{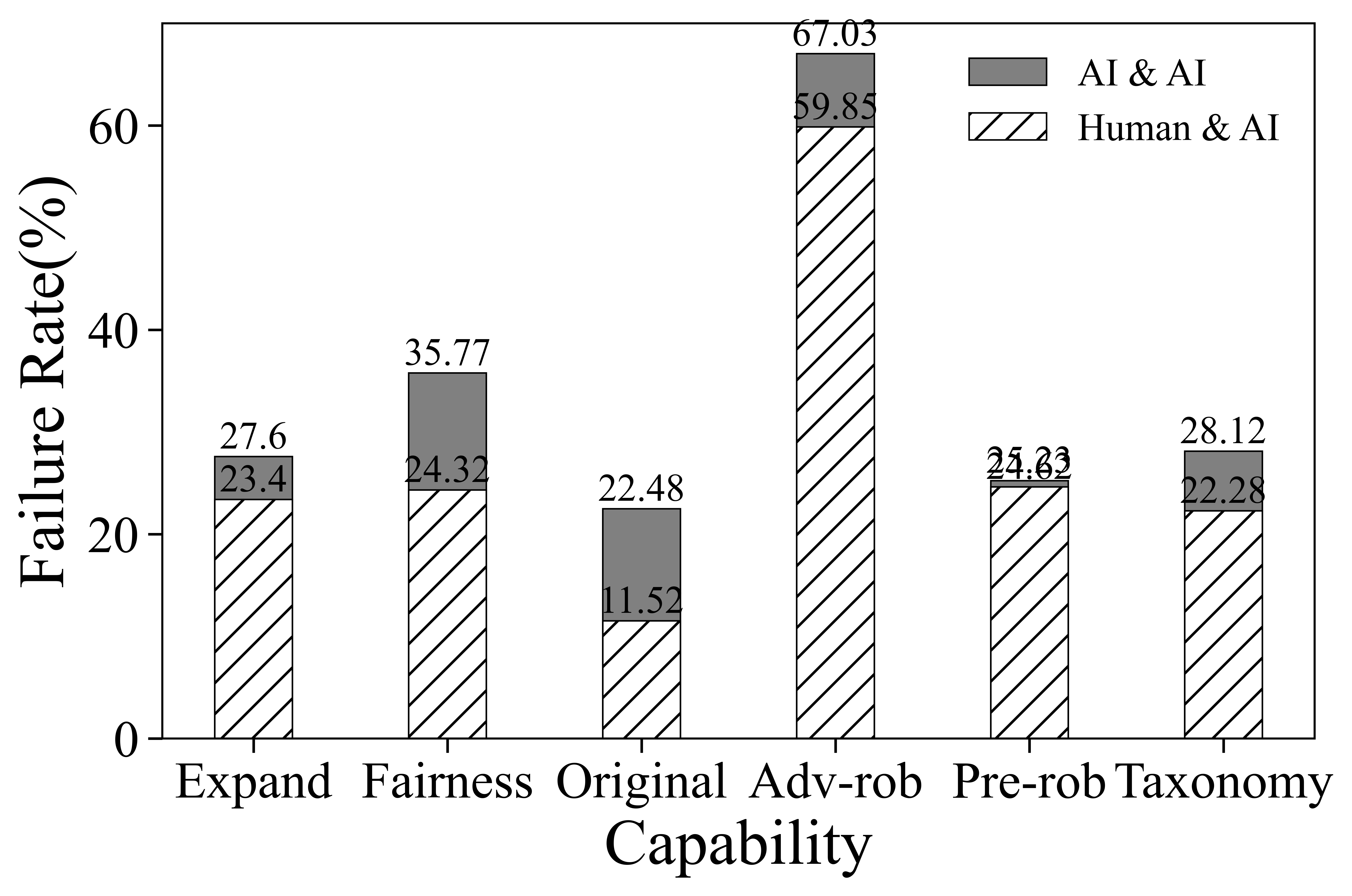}
	}
	\subfigure[Claude-3-Opus]{
		\includegraphics[width=0.3\textwidth]{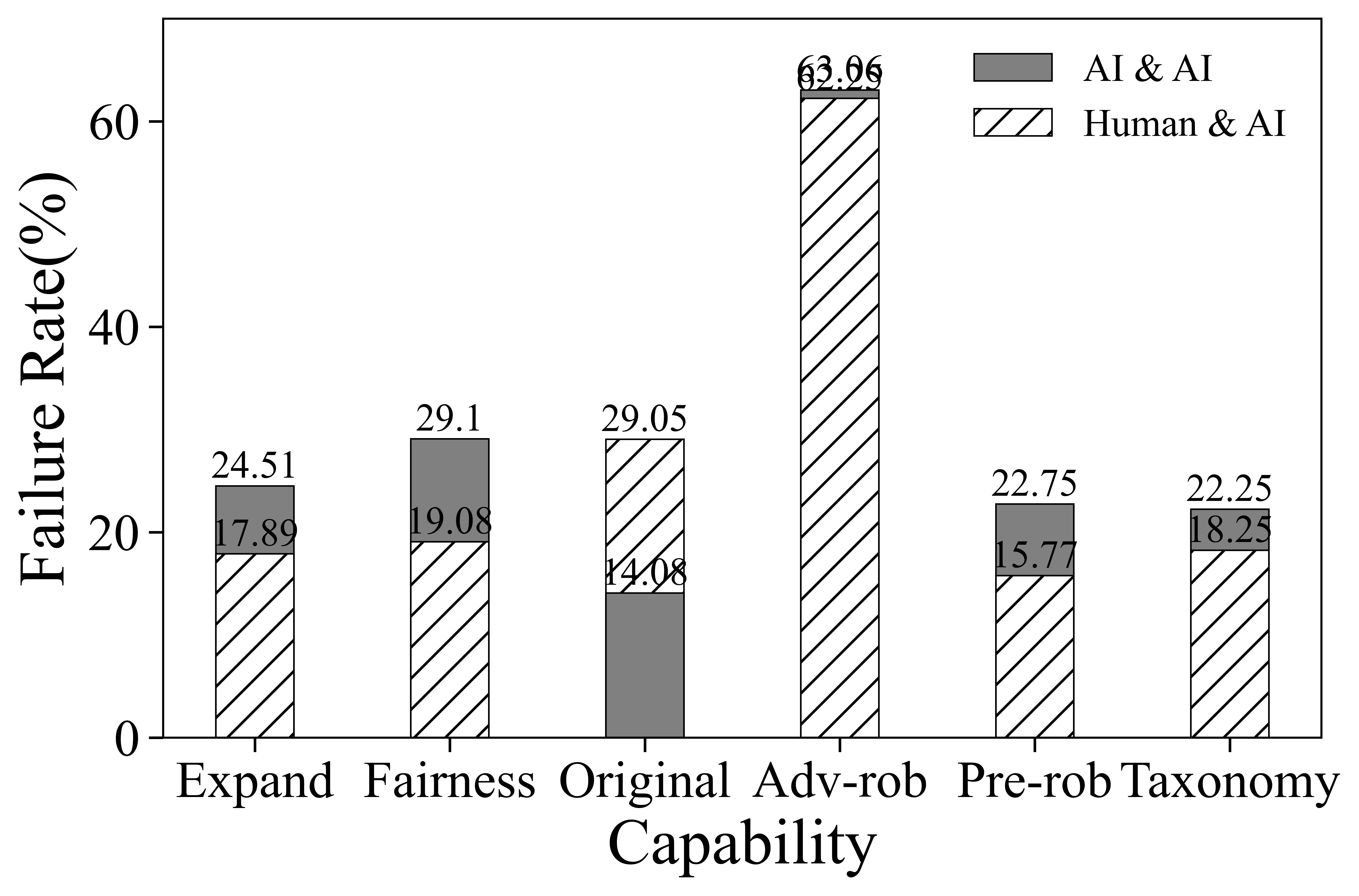}
	}
	\caption{Performance of various capability test suites generated by AutoTestForge using LLMs on Distilbert-base-cased-SST-2.
		The Original and Expand test suites correspond to the transformed test suites in Section \ref{sec:framework-A} by $\mathcal{L}(D_i)$ and $\mathcal{E}(\mathcal{L}(D_i))$, respectively.
		The Fairness, Taxonomy, and Pre-rob test suites refer to the identically named test suites $\mathcal{T}_{fair}$,$\mathcal{T}_{tax}$, and  $\mathcal{T}_{pre-rob}$ in Section \ref{sec:framework-C}, respectively.
		The Adv-rob test suite refers to the suite $\mathcal{T}_{adv-rob}$ from Section \ref{sec:framework-C}.}
	\label{fig:cap}
\end{figure*}

Table~\ref{tab6} shows the performance of AutoTestForge in testing several large-scale models including GPT-3.5-turbo, GPT-4o, LLaMA3-70B, and LLaMA3-8B. Overall, AutoTestForge achieves an average detection rate of $28.94\%$, substantially outperforming CheckList's $17.77\%$. This advantage is evident across different task types: in sentiment analysis tasks, AutoTestForge's detection rate ($25.45\%$) is nearly triple that of CheckList ($9.79\%$), while in semantic textual similarity tasks, AutoTestForge maintains its lead with a $35.42\%$ detection rate over CheckList's $25.74\%$.

Based on the above observations, AutoTestForge shows consistent performance advantages for both large and small models (those smaller than LLMs). The average error detection rate for large models stands at $28.94\%$, while for small models it achieves $32.74\%$. This demonstrates AutoTestForge's ability to maintain stable detection efficiency regardless of model scale. 

Analysis of testing scenarios reveals that the Human \& Human scenario 
consistently achieves higher error detection rates, reaching $42.00\%$ in sentiment analysis tasks and $57.23\%$ in semantic textual similarity tasks. This may be attributed to the fact that LLM-generated test templates tend to align more closely with LLM statistical distributions. In contrast, humans can introduce more non-statistical and creative test cases that better expose potential LLM issues.
Based on this finding, we recommend a layered testing strategy: using the AI \& AI mode for large-scale automated testing during routine testing phases, so as to establish basic test coverage, while incorporating the Human \& Human or Human \& AI modes for critical function verification or quality control checkpoints. This strategy can achieve an optimal balance between cost and effectiveness.

\begin{table}[]
	\caption{Comparison of error detection rates using AutoTestForge across different LLMs.}
	\centering
	\label{tab6}
	\resizebox{\linewidth}{!}{
		\begin{tabular}{ccccccc}
			\toprule
			\multicolumn{2}{c}{Data Source} & GPT-3.5    & GPT-4o           & Llama3-70B       & Llama3-8B        & Avg              \\  \hline
			\multicolumn{7}{c}{Sentiment Analysis Task (SA)}   \\ \hline
			\multirow{2}{*}{Human \& Human} 	& CheckList            & $1.87\% $          & $2.16\% $          & $17.97\%  $        & $18.68\% $         & $10.17\%  $        \\
			& AutoTestForge        & $30.45\%$ 			& $37.38\%$ 		 & $31.50\%$ 		  & $42.00\%$ 		   & $35.33\%$			 \\ \hline
			\multirow{2}{*}{Human \& AI}    	& CheckList            & $2.48\% $          & $7.72\%$           & $4.41\%$           & $30.93\% $         & $11.39\%$          \\
			& AutoTestForge        & $13.44\%$          & $23.21\%$          & $11.87\%$          & $33.70\%$          & $20.55\%$          \\  \hline
			\multirow{2}{*}{AI \& AI}       	& CheckList            & $2.52\%$           & $10.86\%$          & $4.39\%$           & $13.54\%$          & $7.83\%$           \\
			& AutoTestForge        & $14.29\%$          & $24.61\%$          & $13.30\%$          & $29.63\%  $        & $20.46\%$    \\    \hline
			\multirow{2}{*}{Avg} 				& CheckList     & $2.29\%$  & $6.91\%$  & $8.92\%$  & $21.05\%$ & $9.79\%$  \\
			& AutoTestForge & $19.39\%$ & $28.40\%$ & $18.89\%$ & $35.11\%$ & $25.45\%$ \\ \hline
			
			\multicolumn{7}{c}{Semantic Textual Similarity Task (STS)}                 \\ \hline
			\multirow{2}{*}{Human   \& Human} 	& CheckList             & $ 25.16\% $ & $ 6.48\%  $ & $ 28.13\% $ & $ 50.94\% $ & $ 19.92\%$ \\
			& AutoTestForge         & $ 28.31\% $ & $ 19.05\% $ & $ 33.28\% $ & $ 57.23\% $ & $ 26.88\%$ \\  \hline
			\multirow{2}{*}{Human \& AI}      	& CheckList             & $ 32.95\% $ & $ 26.72\% $ & $ 25.86\% $ & $ 32.95\% $ & $ 28.51\%$ \\
			& AutoTestForge         & $ 37.69\% $ & $ 31.45\% $ & $ 29.20\% $ & $ 43.87\% $ & $ 32.78\%$ \\	\hline
			\multirow{2}{*}{AI \& AI}         	& CheckList             & $ 31.37\% $ & $ 27.66\% $ & $ 27.38\% $ & $ 19.91\% $ & $ 28.80\%$ \\
			& AutoTestForge         & $ 40.79\% $ & $ 34.11\% $ & $ 37.92\% $ & $ 40.86\% $ & $ 37.61\%$ \\	\hline
			\multirow{2}{*}{Avg}               	& CheckList             & $ 29.83\% $ & $ 20.29\% $ & $ 27.12\% $ & $ 34.60\% $ & $ 25.74\%$ \\
			& AutoTestForge         & $ 35.60\% $ & $ 28.20\% $ & $ 33.47\% $ & $ 47.32\% $ & $ 32.42\%$ \\	\hline
			
			\multicolumn{7}{c}{Average for SA and STS}   \\ \hline
			\multirow{2}{*}{Avg}              	& CheckList            & $ 16.06\% $ & $ 13.60\% $ & $ 18.02\% $ & $ 27.83\% $ & $ 17.77\%$ \\
			& AutoTestForge        & $ 27.50\% $ & $ 28.30\% $ & $ 26.18\% $ & $ 41.22\% $ & $ 28.94\%$ \\
			
			\bottomrule
	\end{tabular}}
\end{table}

\begin{tcolorbox}[colback=gray!10,colframe=black,arc=2mm, auto outer arc,title={Answer to RQ5},breakable]
	AutoTestForge is effective in evaluating LLMs, consistently outperforming baseline methods with higher error detection rates across both large and small models. 
\end{tcolorbox}


In conclusion, experimental results demonstrate AutoTestForge's superior performance across diverse testing scenarios. Compared to traditional datasets and current state-of-the-art testing tools, AutoTestForge achieves average error detection rates $10-20\%$ higher than existing solutions. Furthermore, AutoTestForge maintains consistent performance across various operational modes - manual guidance, human-AI collaboration, and pure AI - while effectively handling models of different architectures.
Besides, the comparative analysis of test template generation across different large language models reveals task-specific variations in effectiveness. While individual LLMs exhibit distinct strengths in different tasks, 
AutoTestForge's processing methodology effectively closes these performance gaps, producing more consistent results across templates generated by different LLMs.
Another observation regarding template effectiveness is that human-designed test templates exihibit superior performance in detecting LLM errors compared to LLM-generated templates. This phenomenon is probably because LLM-generated templates adhere more closely to the statistical distributions present in LLMs' training data, whereas human-designed templates introduce more diverse, non-statistical test scenarios that better challenge model capabilities.
These findings validate AutoTestForge as a robust testing framework that harmonizes automated and manual testing approaches while maintaining consistent performance improvements across various model architectures and testing scenarios.

\subsection{Discussion}


Bhatt's research \cite{bhattCasestudyefficacy2021} illuminates the inherent efficiency limitation in traditional manual creation of test templates. The finding indicates that an annotator needs approximately one hour to generate $5-7$ templates encompassing $1-2$ capabilities, and in this way $0.5-2$ working days to produce a comprehensive set of multi-dimensional test templates. This significant labor cost and limited output rate severely restrict the expansion of test coverage. In contrast, AutoTestForge utilizes LLMs for automatic test generation, requiring only minimal manual operation of prompt adjustment. This process can be completed at the level of minutes, or even less, achieving a multi-fold increase in efficiency compared to traditional methods. The framework enhances both efficiency and capabilities coverage, autonomously constructing challenging cases across the vocabulary, fairness, and robustness dimensions.

Figure \ref{fig:cap} illustrates the performance of various capability test suites generated by AutoTestForge on the DistilBERT-base-cased-SST-2 model. Generally, the detection rate for the Robustness-attack capability substantially exceeds that of other capabilities, which is higher than $50\%$ in most instances. Conversely, the detection rates for the Expand, Fairness, and Robustness-checklist capabilities are relatively low, while the Original and Taxonomy capabilities exhibit intermediate detection rates. Notice that the errors detected by AutoTestForge provide valuable training data for model improvement: each identified error case represents an opportunity to improve the targeted model. Moreover, the framework's ability to detect diverse error types assists in understanding the intricate relationships between different model capabilities. This insight is particularly valuable for developing targeted improvement strategies that address not only individual weaknesses, but also their potential interactions and compound effects.

Built upon the proven CheckList methodology, AutoTestForge employs a flexible template-based approach adaptable to various NLP classification tasks, such as Natural language inference (NLI) and Named Entity Recognition (NER). The generated test cases can be used for multi-dimensional capability analysis to identify and address model weaknesses, facilitate the creation of targeted improvement strategies, provide high-quality training data for model fine-tuning, support the construction of comprehensive evaluation benchmarks, and enable systematic assessment of model robustness across different dimensions.

\section{Related Work} \label{sec:related}

\subsection{Large Language Models in Software Testing}
Large language models have emerged as a transformative force in software testing, attracting significant research attention for their potential to enhance quality assurance processes.
Recent studies demonstrate that LLMs exhibit exceptional value throughout the software testing lifecycle. During the test preparation phase, LLMs are employed in unit test case generation \cite{2023icsetestcompletion}, test oracle construction \cite{2023icsetestcompletion}, and system test input creation \cite{10172490}. These applications enable early problem detection and prevent issues from progressing into later development stages. In later testing and maintenance phases, LLMs have been proven equally valuable for bug report analysis \cite{WenCheng}, defect classification \cite{icse2023Explaining}, fault localization \cite{icse2024CrashTranslator}, and automated repair tasks \cite{10298532}.

Despite the initial success, the integration of LLMs in software testing continues to face several challenges that necessitate further research and exploration~\cite{li2024largelanguagemodelstest}.
These challenges involve ensuring comprehensive test case diversity and coverage, enhancing the accuracy of test oracles, and achieving seamless integration with established testing methodologies. Addressing these challenges is essential for fully realizing the potential of LLMs in software testing environments.
Researchers have also raised concerns regarding the computational costs and environmental impact of LLMs in software engineering applications \cite{shiEfficient}. They propose a vision for developing efficient and environmentally sustainable LLM4SE, emphasizing the importance of reducing computational resource consumption and environmental impact while maintaining effectiveness in software engineering tasks.

The AutoTestForge framework addresses these challenges from multiple aspects. It employs optimized prompt templates to guide test generation, implements a label validation mechanism for differential testing, and utilizes multi-dimensional expansion techniques. This comprehensive strategy helps overcome the limitations commonly associated with LLM integration in testing workflows while maintaining high testing standards and effectiveness.

\subsection{Behavioral Testing in NLP Models}

Behavioral testing refers to testing different capabilities of a system by verifying input-output behavior, without the need to know the system's internal structure \cite{postonAutomatingspecificationbasedsoftware1996}. The CheckList framework \cite{ribeiroAccuracyBehavioraltesting2020} introduces behavioral testing methods from software engineering into NLP model testing. Specifically, it provides a list of language capabilities applicable to most tasks, guiding users during the testing. At the same time, to break down potential capabilities into specific behaviors, it introduces different types of tests.
The framework has been widely applied in various NLP fields. Junjie Wu et al.~\cite{wuGeneralErrorDiagnosis2023} apply CheckList to the machine translation field and develop the BTPGBT framework.
Paul Röttger et al.~\cite{rottgerHateCheckFunctionalTests2021} develop the HATECHECK tool for testing hate speech detection models.
Betty van Aken et al.~\cite{vanakenWhatYouSee2022a} apply CheckList to the clinical medicine field to evaluate the responses of clinical models to certain changes in input.
The M2C~\cite{hlavnovaEmpoweringCrosslingualBehavioral2023} framework extends CheckList's testing capabilities for specific language features in 12 diverse language types to evaluate the models' generalization ability on type differences in practical applications.

However, the CheckList framework faces various efficiency challenges, including suboptimal resource utilization~\cite{bhattCasestudyefficacy2021}, high complexity of template design, and difficulty in capturing the intricate interactions between different language capabilities. These limitations more or less hinder the comprehensiveness and effectiveness of the testing process. Researchers have proposed several approaches that incorporate language models to alleviate these problems, such as TestAug~\cite{yangTestAugFrameworkAugmenting2022} and SYNTHEVAL~\cite{zhao-etal-2024-syntheval}. However, these methods still heavily rely on human involvement.

Different from these methods, AutoTestForge employs LLMs to provide a fully automated approach that eliminates the need for manual intervention. To address the above challenges, the framework also implements comprehensive multi-dimensional evaluation capabilities for models, encompassing critical assessment dimensions including fairness, robustness, and taxonomy. It provides an automated, comprehensive, and efficient evaluation methodology for NLP models.

\section{Conclusion and Future Work} \label{sec:con}
This paper proposes AutoTestForge, an innovative testing framework to overcome inherent limitations in current NLP model evaluation methods, such as over-reliance on expert knowledge and inability to fully exploit the potential of test cases.
AutoTestForge generates language templates through interaction with LLMs and expands them to produce a vast number of high-quality test samples, thereby broadening the test coverage to capture more potential system defects.
Comprehensive evaluation across sentiment analysis and semantic textual similarity tasks demonstrates that AutoTestForge consistently outperforms existing datasets and testing tools with higher error detection rates ($30.89\%$ for SA and $34.58\%$ for STS on average), with different generation strategies exhibiting comparable effectiveness ($29.03\% - 36.82\%$).

In the near future, we are going to expand AutoTestForge's capabilities by presenting more detailed case studies and conducting comprehensive testing across various NLP tasks. 
Additionally, we will enhance the framework across multiple dimensions. We plan to improve test coverage and testing efficiency while incorporating advanced techniques like active learning, which enables dynamic optimization of testing strategies and iterative identification of potential model defects.


\bibliographystyle{ACM-Reference-Format}
\bibliography{sample-base}

\appendix

\section{Details of the Prompt design} \label{app-A}

\subsection{A prompt used to generate sentence structure descriptions in sentiment analysis task}
\label{app-promt-des}

%

\begin{tcolorbox}[colback=gray!10,colframe=black,arc=2mm, auto outer arc,title={\footnotesize{\textbf{System Prompt}}},breakable]
	\small{
		As a linguist, your expertise is in modifying sentence structures and analyzing emotions. 
		
		You can construct completely different sentence structures based on different emotional analysis tasks. Each sentence structure will be unique and highly representative.
	}
\end{tcolorbox}

\begin{tcolorbox}[colback=gray!10,colframe=black,arc=2mm, auto outer arc,title={{\footnotesize \textbf{Context}}},breakable]
	\small{
		Now I will give you some definitions, please understand and remember:
		\\
		\\
		\#\#\# DEFINITIONS
		
		Description: The sentence refers to the structure of sentences. These sentence structures serve the purpose of emotion analysis tasks, and the sentences generated based on these descriptions can help identify some flaws in the model.
		
		An example is 
		"A positive sentiment sentence with negative sentiment question and word no as the answer."
		\\
		\\
		\#\#\# RETURN
		
		[Description1,Description2,Description3,...]
		\\
		\\
		Your current task is to generate 6 sentence structure descriptions that can be expressed as negative emotions, but these sentence structure descriptions generate sentences that may be incorrectly interpreted as negative emotions by the model task.
		Please ensure that the sentence structures you generate include at least one of the following capabilities: event sequence, negation, anaphora, semantic role labeling, and logic.
		Please note that the sentence will ultimately express a negative emotion.
		Please start with "A  negative  sentiment sentence. " in every description. 
		Not give me other word. Just the list in python format.
	}
\end{tcolorbox}

\subsection{A prompt used to generate templates in sentiment analysis task} \label{app-promt-temp}

\begin{tcolorbox}[colback=gray!10,colframe=black,arc=2mm, auto outer arc,title={\footnotesize{\textbf{System Prompt}}},breakable]
	\small{
		As a linguist and film critic, your expertise is in revising sentence structure and analyzing emotion.
		You can construct completely different sentence structures based on different sentiment analysis tasks. Each sentence structure will be unique and highly representative.
	}
\end{tcolorbox}

\begin{tcolorbox}[colback=gray!10,colframe=black,arc=2mm, auto outer arc,title={\footnotesize{\textbf{Context}}},breakable]
	\small{
		I will give you some definitions, please understand and remember:
		\\
		\\
		\#\# DEFINITIONS
		$1.$ Description: refers to the sentence structure of a sentence. If I want you to generate some sentences based on a description, make sure those sentences have a structure like this.\\
		$2.$ Template: refers to the word-filling template, which is evolved from the sentence structure. The candidate words in the template are wrapped with "{}", and the candidate word set will be put into the pool\\
		$3.$ label: The current task contains $2$ labels, namely $0-$negative, $1-$positive\\
		$4.$ pool: a collection of words that need to be filled with '{}' in the template. And it is a dict in python.\\
		$5.$ Example: Fill in the complete sentence obtained by filling in the template with the word pool.
		\\
		\\
		Here is an example:
		
		\#\# EXAMPLE
		
		"""\\
		Description: Negative sentiment sentences with negative positive words.\\
		Template: {I} {neg\_verb} {thing}.\\
		pool: me: [me, him, she, mary, them], neg\_verb: [hate, dislike], stuff: ["basketball", "ball", "anything"].\\
		For example: I hate everything.\\
		important\_keys: neg\_verb\\
		"""\\
		\\
		\#\#\# RETURN
		Please return it in json format. The format should be:
		\{\\
		\indent\enspace``Description": \verb|<term>|, \\
		\indent\enspace``Templates":[ \\
		\indent\quad\enspace\{\\
		\indent\qquad\enspace``template":\verb|<term>|,	\quad	``label":\verb|<term>|, \\
		\indent\qquad\enspace``pool":\verb|<term>|,		\qquad \enspace	``example": \verb|<term>|, \\
		\indent\qquad\enspace``check\_label":\verb|<term>|,\quad \enspace	``score":\verb|<term>| \\
		\indent\quad\enspace\},\\
		\indent\quad\enspace\{\verb|<term>|\},***] \\
		\}\\
		\\
		Now I will give you some sentence structure descriptions, and ask you to generate corresponding templates, candidate words, and tags based on these descriptions. Each sentence description requires $3$ templates.\\
		\\
		Descriptions:\\
		$1.$ \verb|<term>|\\
		\\
		One thing you need attention is that this description all express some negative emotion. So please you according this descriptions give me some template can express negative emotion.\\
		Please ensure that the templates and sentences are natural. Sentences with a score of $9.5$ or above out of $10$ will be used.\\
		Please make sure that the given template can find the defects of the model.\\
		Please note that when words are repeated, you can use other ways to avoid them, but please do not end with numbers\\
		\\
		Please generate some templates about movie reviews.\\
		Return json file, Not other format.  \\
		Attetntion: Must express negative emotion! negative emotion (The label is $0$)! negative emotion(The label is $0$)! negative emotion(The label is $0$)!
	}
\end{tcolorbox}

\subsection{The prompt in experiment \ref{sec-q-5} } \label{app-prrompt-q5}

\begin{tcolorbox}[colback=gray!10,colframe=black,arc=2mm, auto outer arc,title={\footnotesize{\textbf{A prompt for SA questions}}},breakable]
	\small{
		Analyze the sentiment of the text enclosed in square brackets, determine if it is positive, or negative, and return the answer as the corresponding sentiment label "positive-1" or "neutral-0".\\
		$[\{\tt{text}\}]$ = \\
		Please reply in this form: Ans=negative-0/positive-1\\
	}
\end{tcolorbox}

\begin{tcolorbox}[colback=gray!10,colframe=black,arc=2mm, auto outer arc,title={\footnotesize{\textbf{A prompt for STS questions}}},breakable]
	\small{
		Task: Determine whether the following two sentences are semantically similar. This is a binary classification task where 1 indicates semantic similarity and 0 indicates semantic dissimilarity.\\
		
		Sentence 1: $\{\tt{text}\}$\\
		Sentence 2: $\{\tt{text}\}$\\
		
		Please analyze the core semantic meaning of these two sentences and provide your classification (1 or 0).
		Please reply in this form: Ans=dissimilarity-0/similarity-1.
	}
\end{tcolorbox}

\end{document}